\documentclass[twocolumn]{aastex62}

\usepackage{natbib}
\usepackage{color}
\usepackage{graphicx}
\usepackage[caption=false]{subfig}
\usepackage{multirow}

\received{July 24, 2020}
\revised{September 04, 2020}
\accepted{September 11, 2020}
\submitjournal{ApJ}

\shorttitle{Surface Brightness Evolution}
\shortauthors{A. Whitney}

\begin{document}

\title{Surface Brightness Evolution of Galaxies in the CANDELS GOODS Fields up to $z \sim 6$: High-z Galaxies are Unique or Remain Undetected}

\author{A. Whitney}
\affil{University of Nottingham, School of Physics \& Astronomy, Nottingham, NG7 2RD, UK}

\author{C. J. Conselice}
\affiliation{University of Nottingham, School of Physics \& Astronomy, Nottingham, NG7 2RD, UK}
\affiliation{Jodrell Bank Centre for Astrophysics, University of Manchester, Oxford Road, Manchester UK}
\author{K. Duncan}
\affiliation{SUPA, Institute for Astronomy, Royal Observatory, Blackford Hill, Edinburgh, EH9 3HJ, UK}
\affiliation{Leiden Observatory, Leiden University, PO Box 9513, NL-2300 RA Leiden, The Netherlands}

\author{L. R. Spitler}
\affiliation{Research Centre for Astronomy, Astrophysics \& Astrophotonics, Macqaurie University, Sydney, NSW, Australia}




\begin{abstract}

We investigate the rest-frame Ultraviolet (UV, $\lambda\sim2000$\AA) surface brightness (SB) evolution of galaxies up to $z\sim6$ using a variety of deep Hubble Space Telescope imaging. UV SB is a measure of the density of emission from mostly young stars and correlates with an unknown combination of star formation rate, initial mass function, cold gas mass density, dust attenuation, and the size evolution of galaxies. In addition to physical effects, the SB is, unlike magnitude, a more direct way in which a galaxy's detectability is determined.  We find a very strong evolution in the intrinsic SB distribution which declines as $(1+z)^{3}$, decreasing by 4-5 mag arcsec$^{-2}$ between $z=6$ to $z=1$. This change is much larger than expected in terms of the evolution in UV luminosity, sizes or dust extinction and we demonstrate that this evolution is 'unnatural' and due to selection biases. We also find no strong correlation between mass and UV SB. Thus, deep HST imaging is unable to discover all of the most massive galaxies in the distant universe. Through simulations we show that only $\sim15$\% of galaxies that we can detect at $z=2$ would be detected at high-$z$. We furthermore explore possible origins of high SB galaxies at high-$z$ by investigating the relationship between intrinsic SB and star formation rates. We conclude that ultra-high SB galaxies are produced by very gas rich dense galaxies which are in a unique phase of evolution, possibly produced by mergers. Analogues of such galaxies do not exist in the relatively nearby universe.

\vspace{10mm}

\end{abstract}



\section{Introduction} \label{sec:intro}

When studying and examining galaxy evolution as a function of redshift it is common to observe and characterise how quantities evolve with redshift to infer evolution.  These quantities are most famously the UV luminosity function (LF) \citep[e.g.][]{arnouts05, mclure13, bouwens15}, the stellar mass function \citep{duncan14, bhatawdekar19}, the spectral shape of the UV light distribution \citep[e.g.][]{mclure18}, morphological evolution \citep[e.g.][]{conselice09} as well as the merger and pair fraction evolution \citep[e.g.][]{duncan19}.  

Galaxy surface brightness has previously been studied in the optical and near-infrared at low redshifts \citep{schade95, roche98, labbe03, barden05}. Such studies find mixed results with some suggesting there is evolution in the surface brightness \citep[e.g.][]{schade96} and others suggesting it does not exist or is due to selection effects \citep[e.g.][]{simard99}. Higher redshift galaxies have been examined in the context of gas, star formation, and star surface density but rarely explicitly in the form of surface brightness for the last decade or so.  However these density relations are similar to the SB problem. Since the low redshift studies, larger and deeper surveys have taken place allowing us to now look further and deeper into the evolution of surface brightness of galaxies, and to do so at a consistent rest-frame wavelength.   With modern deep imaging data from the Hubble Space Telescope we can investigate the evolution of surface brightness as measured in the ultraviolet (UV) and determine how it evolves in terms of intrinsic surface brightness, and what this evolution implies for galaxy formation, star formation, dust as well as galaxy detection.  Surface brightness in the UV rest-frame is a good indicator of star formation density and gas density \citep{schmidt59, kennicutt98, leroy08, freundlich13}, and thus allows us to understand how these quantities change with time to first order. Star formation density has been shown to increase to a peak between $z \sim 1$ and $z \sim 2$, before becoming nearly constant to high redshift for Lyman break galaxies \citep[e.g.][]{steidel99, giavalisco04}. Observations from the \textit{Hubble Space Telescope} show a small amount of evolution at $z > 3$ \citep{bouwens03} and those measurements that are made using photometric redshifts show constant star formation up to $z \sim 6$ \citep{thompson01, kashikawa03}. Surface brightness also determines how well galaxies can be detected and in this paper we investigate both of these problems.

When measuring the surface brightness of a given galaxy it is now well known that this value is constant for the same galaxies seen at different distances in the nearby universe. However, when a galaxy is at cosmological distances, the  surface brightness becomes lower by a significant amount that scales with redshift ($z$) as $(1+z)^{\alpha}$, where $\alpha$ can range from 3 to 5 depending on the particular circumstances. One result of this is that the observability of galaxies which would normally be detectable would have such a low surface brightness that even with deep Hubble Space Telescope exposures we would not be able to detect them at higher redshifts.

The literature on the evolution of SB is however not entirely consistent on how this evolution occurs.  Previous studies of the rest-frame surface brightness evolution of disk galaxies find mixed results ranging from little or no evolution to a difference of $\sim$1-2 mag between the local Universe and $z\sim$ 1. For example, \cite{labbe03} use ground-based near-infrared imaging and find an increase of 1 mag out to $z \sim 2-3$ for six Milky Way-type galaxies. \cite{schade96} also use ground-based imaging but find a stronger evolution of 1.6 mag from $z \sim 1$ for a larger sample of 143 galaxies (33 early-type and 110 late-type). These results are also found in \textit{Hubble Space Telescope} observations; \cite{schade95}, \cite{lilly98}, \cite{roche98} and \cite{barden05} who find an average increase in surface brightness of $\sim 1$ mag by $z \sim 1$ for a mixture of both spiral and elliptical galaxies. 

In terms of surface brightness and selection, \cite{simard99} identify the need for selection effects to be taken into account when probing higher redshifts, and this is another reason why studying the evolution of SB is important. Before taking selection effects into account, \cite{simard99} find a change of 1.3 mag from $z\sim1$ to $z\sim0$ but once these effects are considered, they find no evolution in the surface brightness. Similarly, \cite{ravindranath04} find a change in surface brightness of $<0.4$ mag over the range $0.2 < z < 1.25$ in the z-band for disk like galaxies when considering selection effects. These studies show that there are relatively bright galaxies in the highest redshift bins. \cite{trujillo04} also take selection effects into account in their analysis, however they find a change of $\sim0.8$ mag from $z = 0.7$ to $z = 0$ when measuring the surface brightness in the V-band. Models supporting an increase in surface brightness include \cite{bouwens02} who model disk evolution based on two different approaches. Both models find a strong evolution in the B-band surface brightness of 1.5 mag by $z = 1$, and they argue that these results are not an artefact of selection effects.

These works are limited to relatively low redshift objects, where the cosmological SB dimming is not as dramatic as it is at higher redshifts, and are also limited to the optical rest-frame.  As we now have access to higher redshift data from surveys such as the Cosmic Assembly Near-infrared Deep Extragalactic Survey (CANDELS) we can now probe how the surface brightness of galaxies changes with time, and infer the processes that cause this evolution over $\sim 12$ Gyr of cosmic time for the UV rest-frame. 

We examine the evolution of the UV rest-frame ($\lambda \sim 2000$\AA) surface brightness through the redshift range of $0.5 < z < 6.5$. We also examine the relationship between the surface brightness with star formation rate (SFR). In order to reduce any biases in the results, we use a mass-selected sample that is $>$95\% complete \citep{duncan19}. We also examine a separate number density-selected sample. Stellar mass incompleteness is an important factor to consider when examining galaxies at high redshift as due to detection limits, some galaxies will be missed in observations. We explore this incompleteness in our sample in order to see how this affects our surface brightness measurements.

This paper is structured as follows. In \S \ref{sec:data} we describe the data and the sample. In \S \ref{sec:method} we describe the method used. In \S \ref{sec:results} we present our main results, and in \S \ref{sec:discussion} we discuss our findings. Finally, we conclude in \S \ref{sec:summ}. Throughout this paper we use AB magnitudes and assume a $\Lambda$CDM cosmology with H$_0$ = 70 kms$^{-1}$Mpc$^{-1}$, $\Omega_m$ = 0.3, and $\Omega_{\Lambda}$ = 0.7.

\section{Data and Sample Selection} \label{sec:data}

For this work two samples of galaxies over the redshift range $0.5 < z < 6.5$ are selected from the Great Observatories Origins Deep Survey North and South (GOODS-N and GOODS-S respectively) fields \citep{giavalisco04} from the Cosmic Assembly Near-infrared Deep Extragalactic Survey (CANDELS) \citep{grogin11, koekemoer11}. CANDELS is a Multi Cycle Treasury Programme which images objects using the Advanced Camera for Surveys (ACS) and the Wide Field Camera 3 (WFC3). In total, CANDELS covers an area of 800 arcmin$^2$ over five different fields. The GOODS-N and GOODS-S fields cover a combined area of 160 arcmin$^2$ and are centred on the Hubble Deep Field North and the Chandra Deep Field South \citep{giavalisco04}. CANDELS is divided into CANDELS/Deep which images both GOODS-N and GOODS-S to a 5$\sigma$ depth of 27.7 mag in the H-band and CANDELS/Wide which images the fields to a 5$\sigma$ depth of 26.3 mag in the H-band. For this work, we use the F435W, F606W, F814W, and F850LP images observed using the ACS and the F105W, F125W, and F160W images observed using the WFC3. These filters will be referred to as B$_{435}$, V$_{606}$, I$_{814}$, z$_{850}$, Y$_{105}$, J$_{125}$, and H$_{160}$ from this point onwards. 

\begin{figure*}
\centering
\begin{tabular}{cc}
\subfloat{\includegraphics[width = 0.475\textwidth]{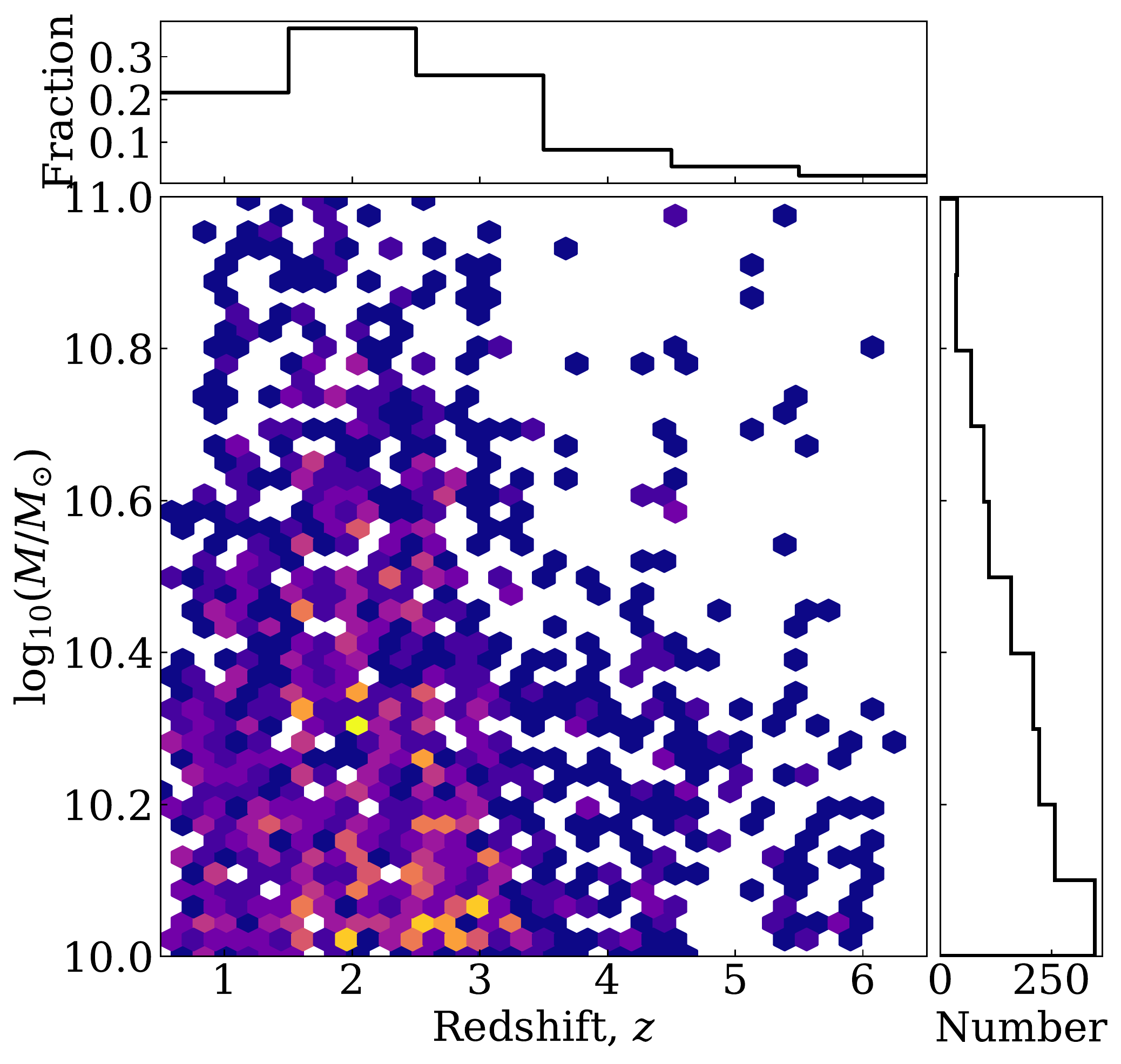}} &
\subfloat{\includegraphics[width = 0.475\textwidth]{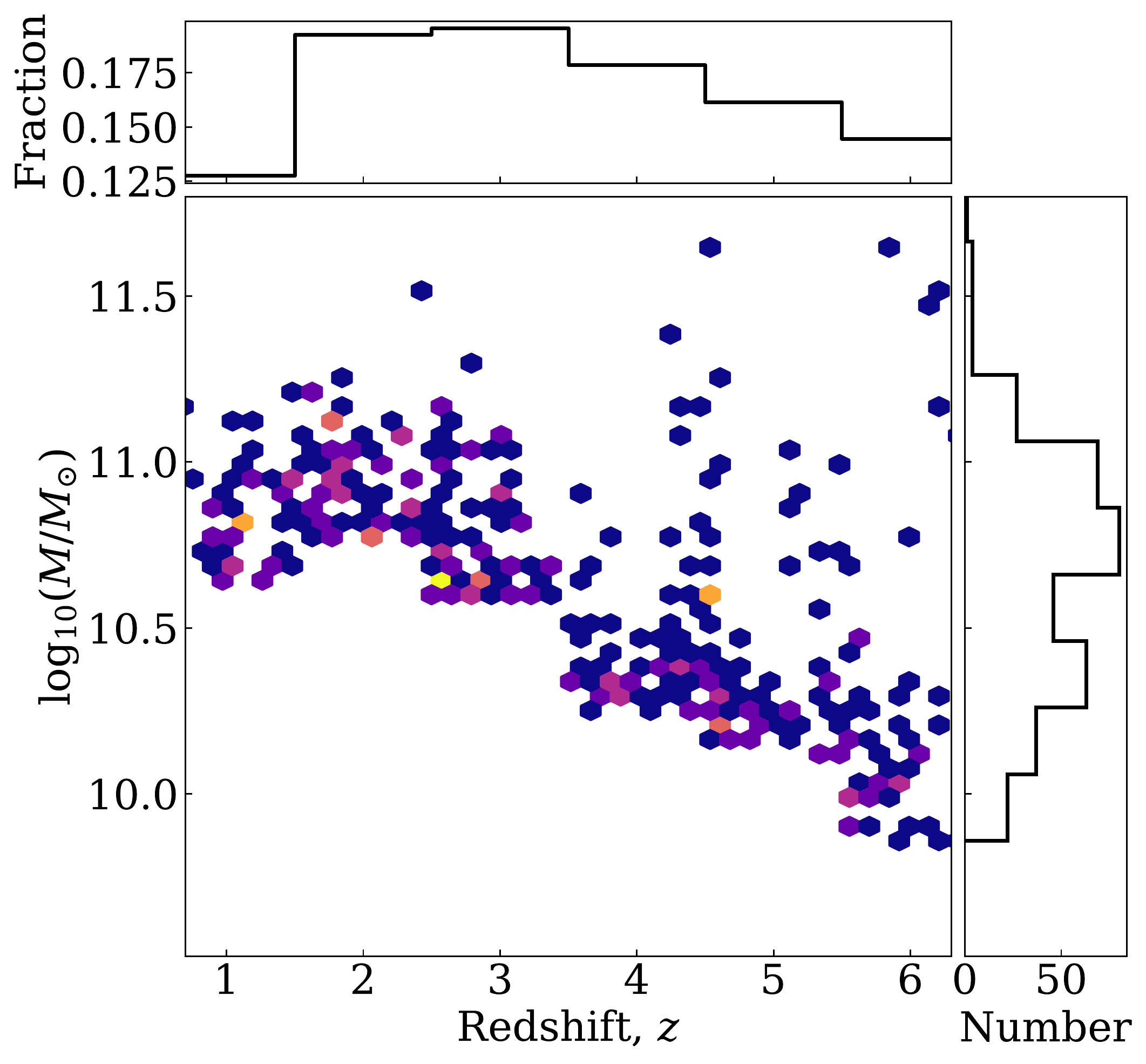}} \\
\end{tabular}
\caption{Left: Redshift and mass distributions of the mass-selected sample. Right: Redshift and mass distributions of the number density-selected sample. In both, yellow indicates an increased density of points and purple indicates a low density. The colour scaling is linear.}
\label{fig:zdist}
\end{figure*}

In this work, we analyse two separate samples; a mass-selected sample and a number-density selected sample. The mass-selected sample is comprised of 1,522 galaxies that lie within the mass range $10^{10}M_{\odot} < M_* < 10^{11}M_{\odot}$ to ensure completeness \citep{duncan19}. The completeness limits at $z = 6$ are log$_{10}(M_{\odot})$ = 10. The mass limits for $z < 6$ are considerably lower than our chosen mass limit of $10^{10}M_{\odot}$. The redshift and mass distributions of the galaxies within this sample are shown in the left panel of Figure \ref{fig:zdist}. The yellow regions indicate a high density of galaxies and purple indicates a lower density of galaxies. The colour scaling is linear.

The number density-selected sample is generated by using a constant number density of 1$\times$10$^{-4}$ Mpc$^{-3}$ \citep{ownsworth16}. This sample consists of 400 galaxies that have masses in the range $10^{9.5} M_{\odot} < M_* < 10^{11.8}M_{\odot}$. The redshift and mass distributions within the number density-selected sample are shown in the right panel of Figure \ref{fig:zdist}. As with the left panel, the yellow regions indicate a high density of galaxies and purple indicates a lower density of galaxies and the colour scaling is linear. This sample is chosen as it potentially allows us to directly track the progenitors and descendants of massive galaxies at all redshifts \citep[e.g.][]{mundy15, ownsworth16}. In the following subsections we discussed how we measured photometric redshifts and stellar masses for  our sample of galaxies. There are a small number of high-$z$ galaxies with high masses in this sample that are unlikely to be real however these galaxies do not affect our results.

\subsection{Photometric Redshifts} \label{sec:photoz}

We use the method as described in \cite{duncan19} to calculate photometric redshifts for the galaxies within our samples. Template-fitting estimates are determined using the \textsc{eazy} photometric redshift software \citep{brammer08}. Three separate template sets are used and fit to all available photometric bands. These template sets include zero-point offsets to the input fluxes and additional wavelength-dependent errors. We then calculate further empirical estimates using a Gaussian process code (GPz; \cite{almosallam16}) using a subset of the available photometric bands. Individual redshift posteriors are calibrated and the four estimates are combined in a statistical framework via a hierachical Bayesian combination to produce a final redshift estimate. For further details of the process, see section 2.4 of \cite{duncan19}.

\subsection{Stellar Mass Fitting} \label{sec:mass}

The galaxy stellar masses we use are measured by using a modified version of the SED code described in \cite{duncan14}. The stellar mass is estimated at all redshifts in the photo-$z$ fitting range as opposed to finding the best-fit mass for a fixed input redshift. Also included in these estimates is a so-called "template error function" to account for uncertainties introduced by the limited template set and any wavelength effects. The method for this error function is outlined in \cite{brammer08}. This mass-fitting technique uses \cite{bruzual03} templates and includes a wide range of stellar population parameters and assumes a \cite{chabrier03} initial mass function. The assumed star formation histories follow exponential $\tau$-models for both positive and negative values of $\tau$. Characteristic timescales of $\left|\tau\right|$ = 0.25, 0.5, 1, 2.5, 5, and 10 along with a short burst ($\tau$ = 0.05) and continuous star formation models ($\tau \gg$ 1/$H_0$) were assumed.

We compare the mass measurements we make to the average of those determined by the several teams within the CANDELS collaboration \citep{santini15}. This is done in order to ensure that the stellar mass estimates do not suffer from systematic biases. There is some scatter between the two mass estimates however our mass estimates are not affected by significant biases. For further details on the method and models used, see section 2.5 of \cite{duncan19} for an extensive discussion.

\subsection{Star Formation Rates}

The star formation rates we use in this paper are determined through spectral energy distribution (SED) fitting. The measured rest-frame absolute magnitudes (M$_{1500}$) are corrected for dust extinction using the relation determined by \cite{meurer99}. Where the relation implies a negative extinction, the extinction is set to 0. The UV star formation rates are calculated using the following:

\begin{equation}
    SFR\ (M_{\odot}\ yr^{-1}) = \frac{L_{UV}\ (erg\ s^{-1}\ Hz^{-1})}{1.39 \times 10^{27}}
\end{equation}

\noindent where the conversion factor of \cite{madau98} and \cite{kennicutt98} is used. For further details, see \cite{duncan14} and \cite{duncan19}.

\section{Methodology} \label{sec:method}

\subsection{2D Lyman-break Imaging}

We utilise the image processing technique described in \cite{whitney19} in order to produce the images we ultimately use within our analysis. However, we normally only use imaging in the bands corresponding to the UV rest-frame as opposed to the optical as is done in \cite{whitney19}. The bands used are given in Table \ref{tab:imgprocessing}, along with the rest-frame wavelength we probe at each redshift. This 2-D Lyman break imaging method removes all nearby galaxies and only retains the original system. We do this to avoid contamination from foreground objects, utilising the fact that the Lyman break allows us to find which systems are at a different redshifts of the principal galaxy we are looking at.

The process makes use of the Lyman-break at 912{\AA} and removes contaminating foreground objects from 100 $\times$ 100 pixel postage stamp images of galaxies. This is achieved by subtracting the band corresponding to the Lyman-break from the band corresponding to the UV rest-frame and normalising the resulting image by the UV rest-frame image. Maps of the pixels are created such that the pixels corresponding to the central object are given a value of one, and the pixels corresponding to the sky are given a value of zero. The pixels corresponding to the central object are identified by selecting those that have a value that is greater than or equal to three times the standard deviation of the background statistics. We use this map, along with the segmentation map of the UV rest-frame, to remove the field objects. These removed areas are then replaced with noise that has the same mean and standard deviation as the sky. The process can be described by the equation 

\begin{equation}
    O_{i,j}^{analysis} = \left(\frac{O_{i,j}^{raw}-D_{i,j}^{raw}}{O_{i,j}^{raw}} \cdot S_{i,j}\right) + f(O_{i,j}^{raw, sky}) 
    \label{eq:imgproc}
\end{equation}

\noindent where $O_{i,j}^{raw}$ is the original optical rest-frame image or its substitute, $D_{i,j}^{raw}$ is the original drop-out image, $S_{i,j}$ is the segmentation map, and $f(O_{i,j}^{raw, sky})$ is some function of the raw optical rest-frame image. The function $f(O_{i,j}^{raw, sky})$ creates an image in which the pixels corresponding to the central object are 0, the pixels corresponding to the sky are those of the raw optical rest-frame image, and the pixels corresponding to the field objects are noise that has the same mean and standard deviation of the sky.   Note that we are unable to use this technique at $z < 3$. This is due to the fact that we are unable to observe the Lyman-break for these galaxies, however there are much fewer foreground galaxies at these redshifts and thus contamination is not a serious concern.  For further details and example images, see \cite{whitney19}. 

\begin{figure*}
\centering
\begin{tabular}{cccc}
\LARGE{$D_{i,j}^{raw}$} & \LARGE{$O_{i,j}^{raw}$} & \LARGE{$S_{i,j}$} & \LARGE{$O_{i,j}^{analysis}$} \\
\subfloat{\includegraphics[width = .23\textwidth]{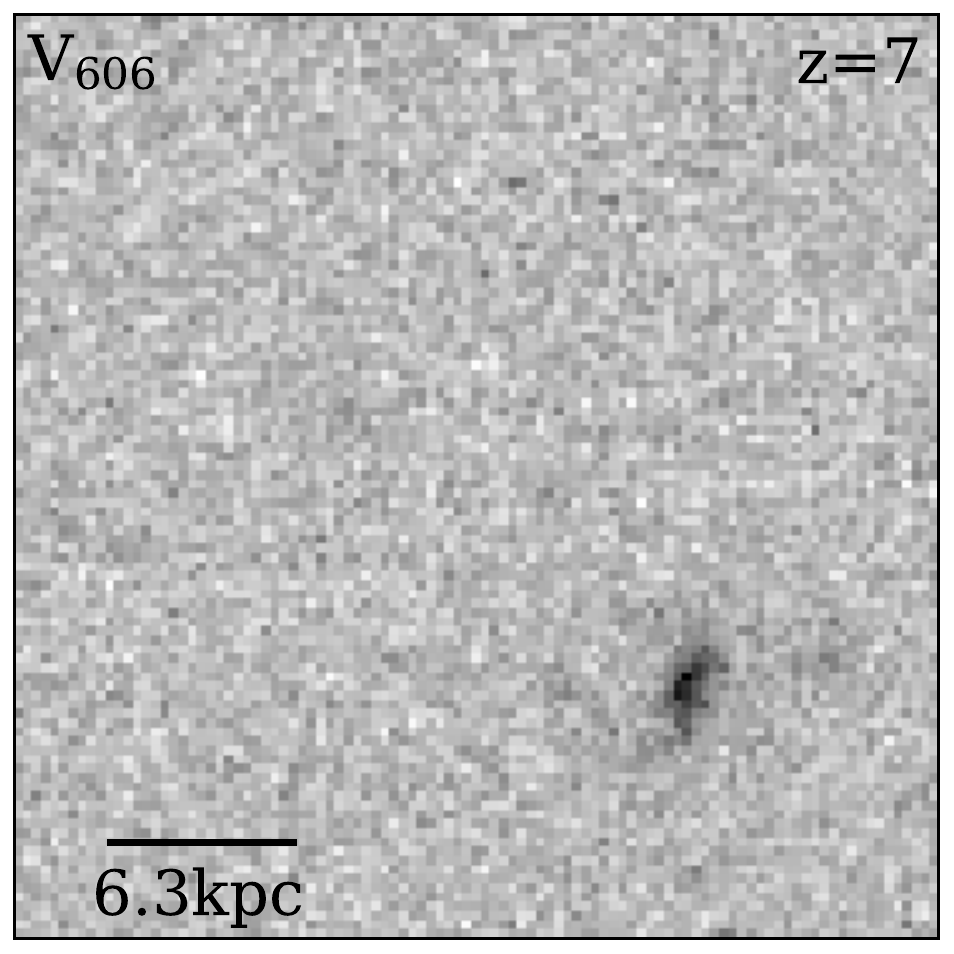}} &
\subfloat{\includegraphics[width = .23\textwidth]{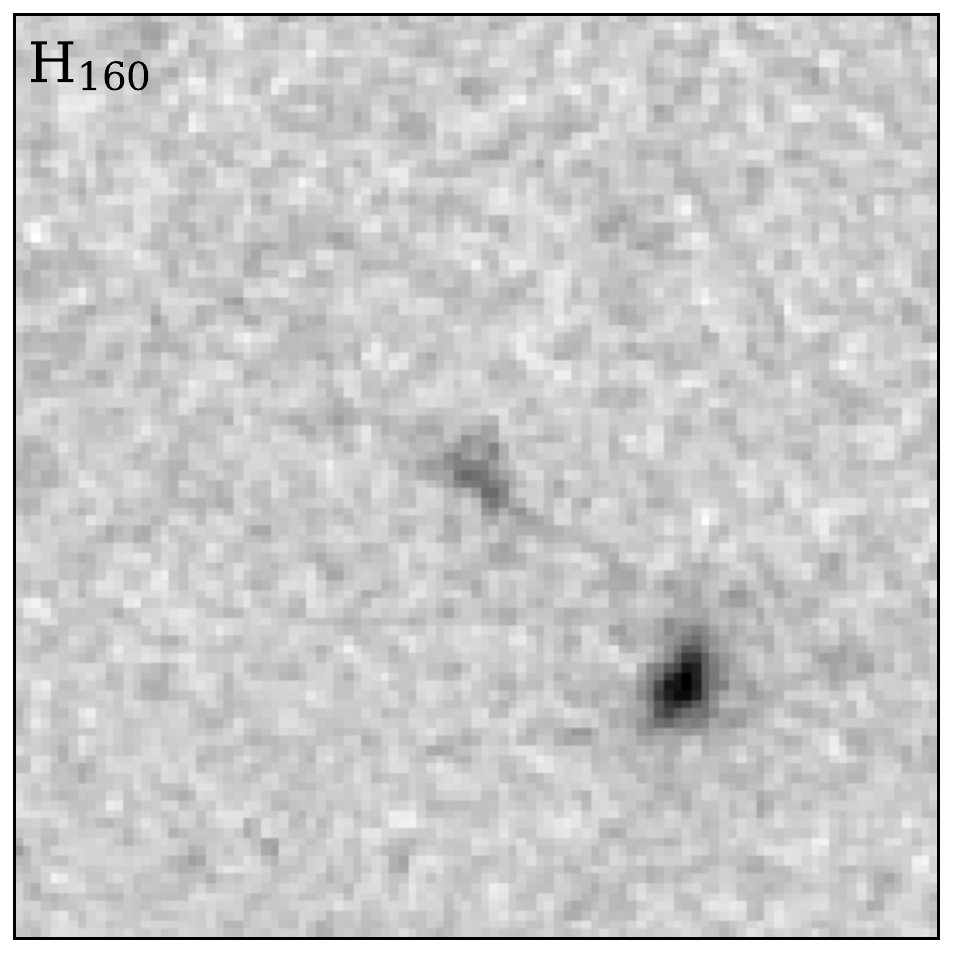}} &
\subfloat{\includegraphics[width = .23\textwidth]{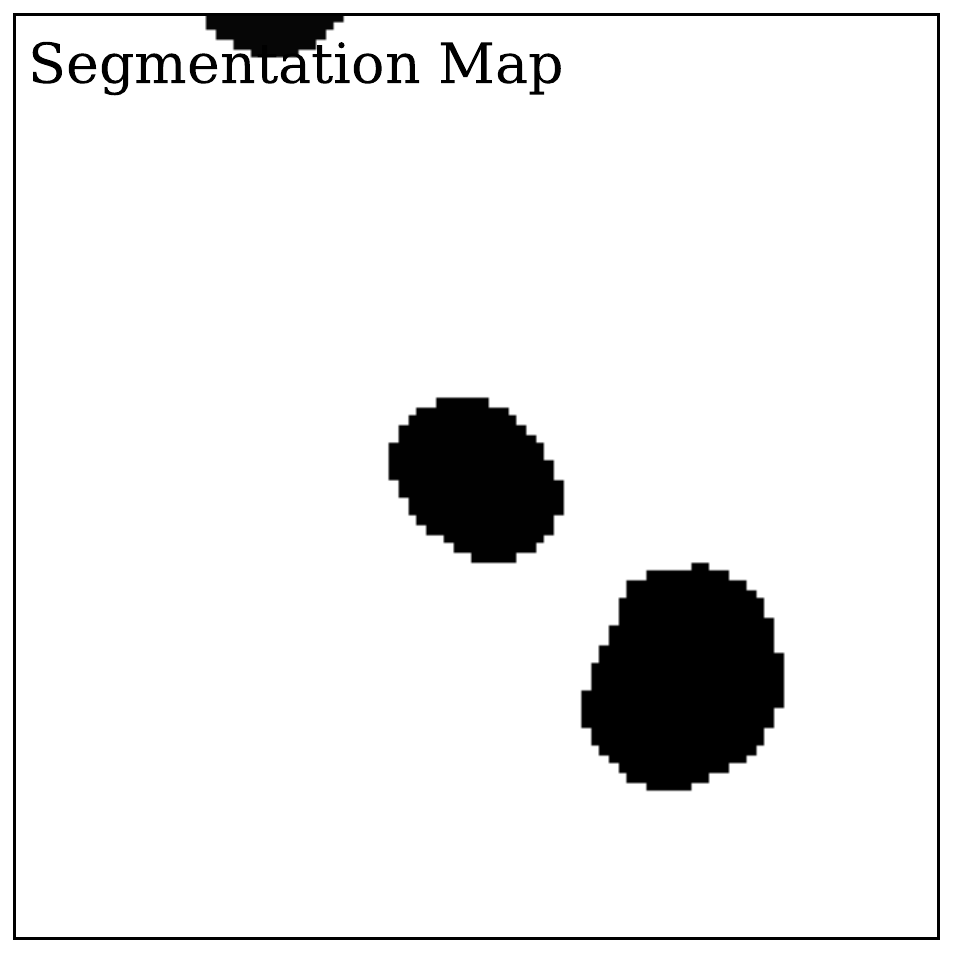}} &
\subfloat{\includegraphics[width = .23\textwidth]{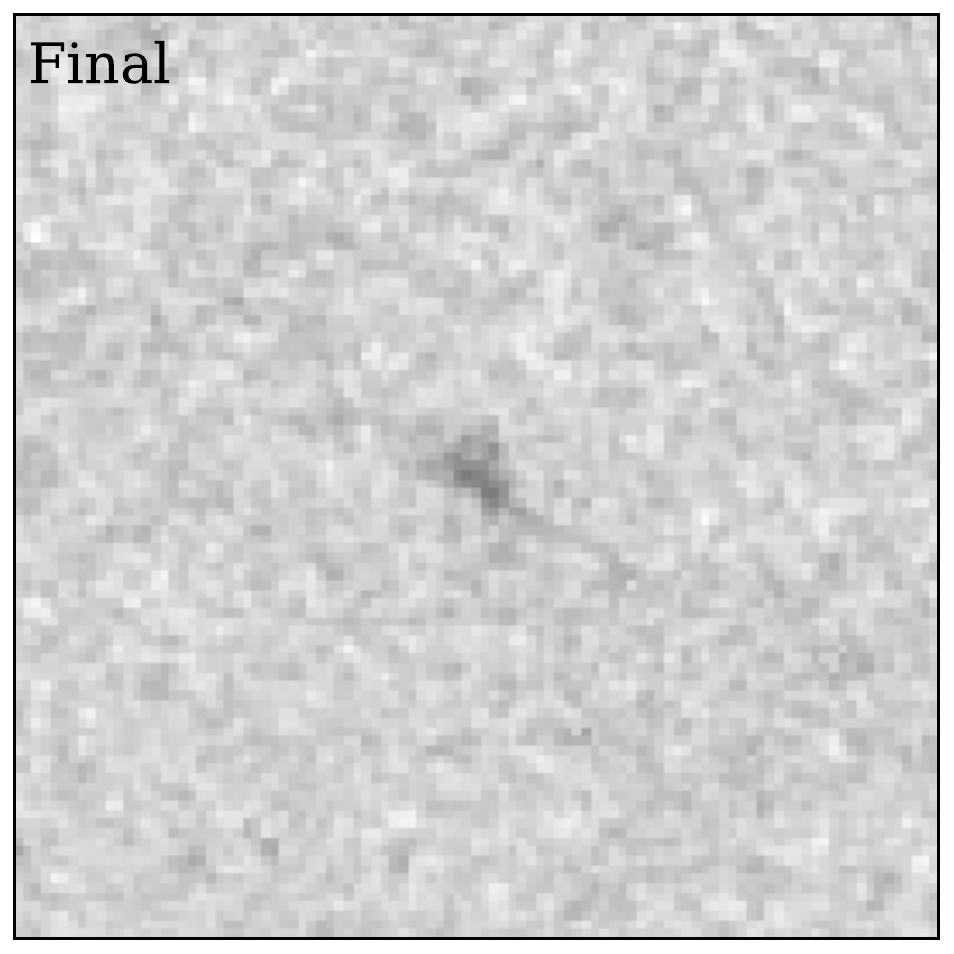}} \\
\subfloat{\includegraphics[width = .23\textwidth]{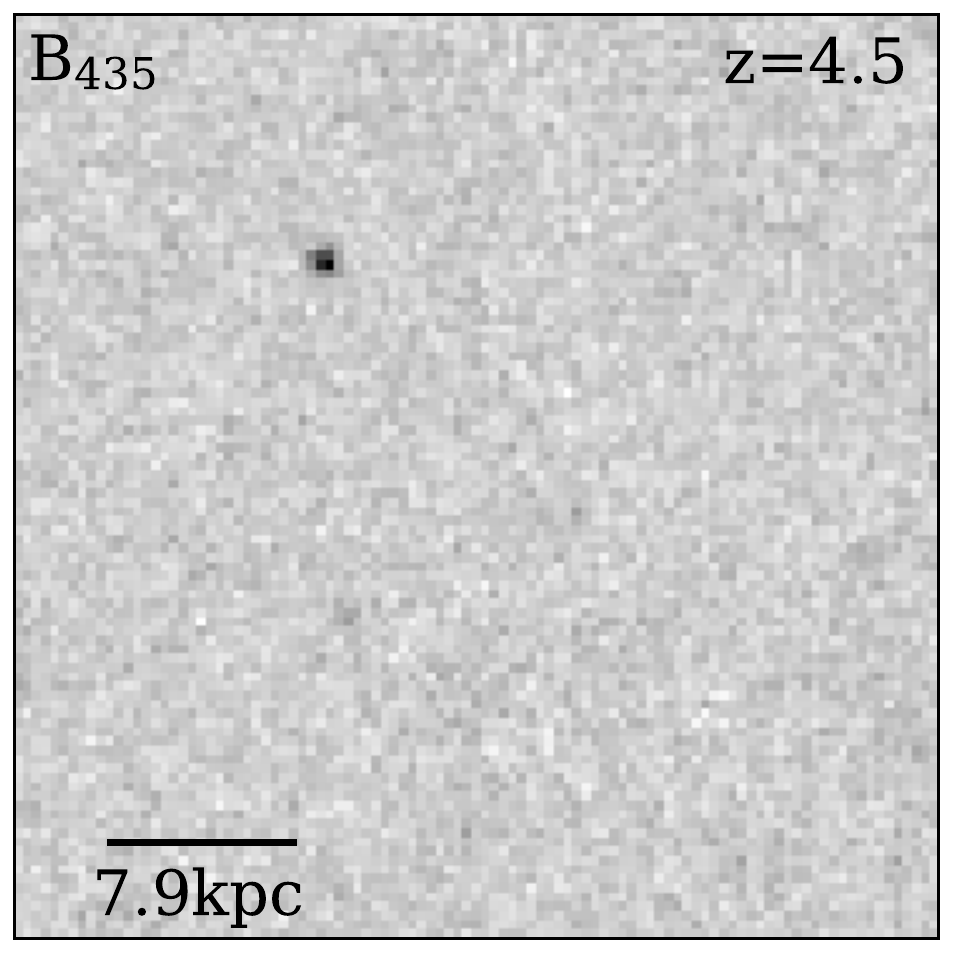}} &
\subfloat{\includegraphics[width = .23\textwidth]{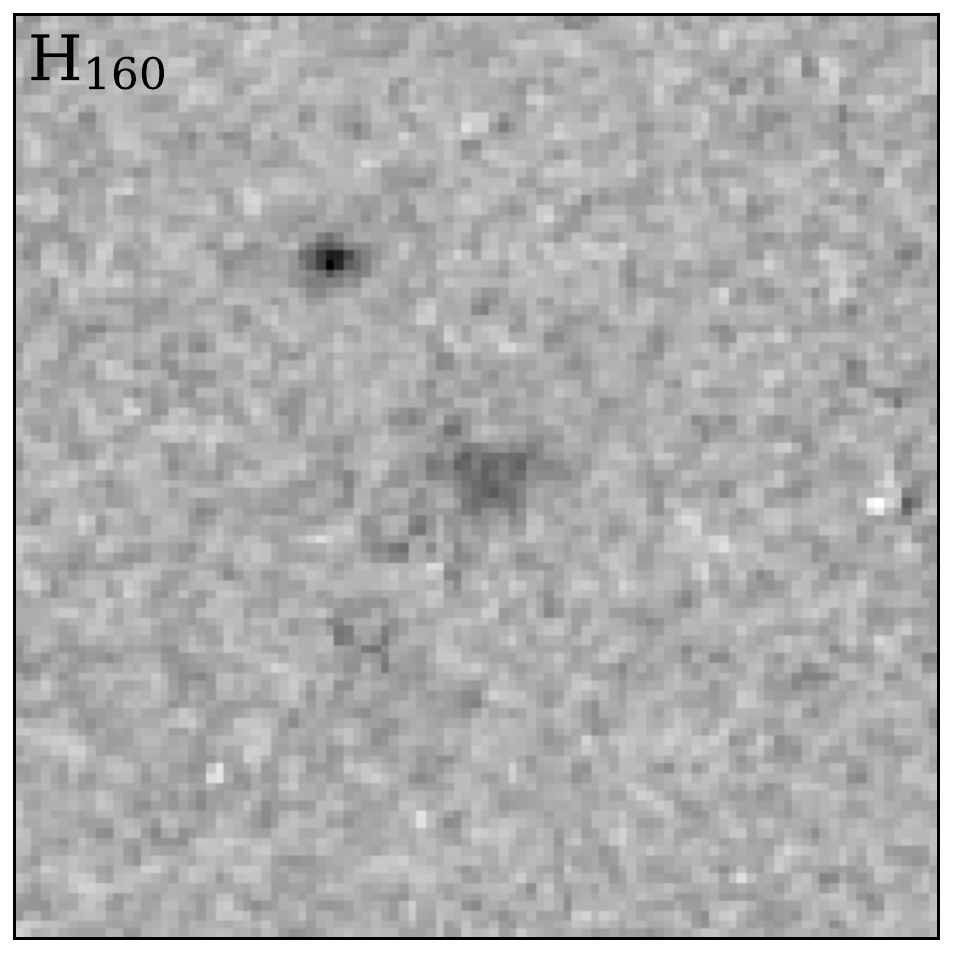}} &
\subfloat{\includegraphics[width = .23\textwidth]{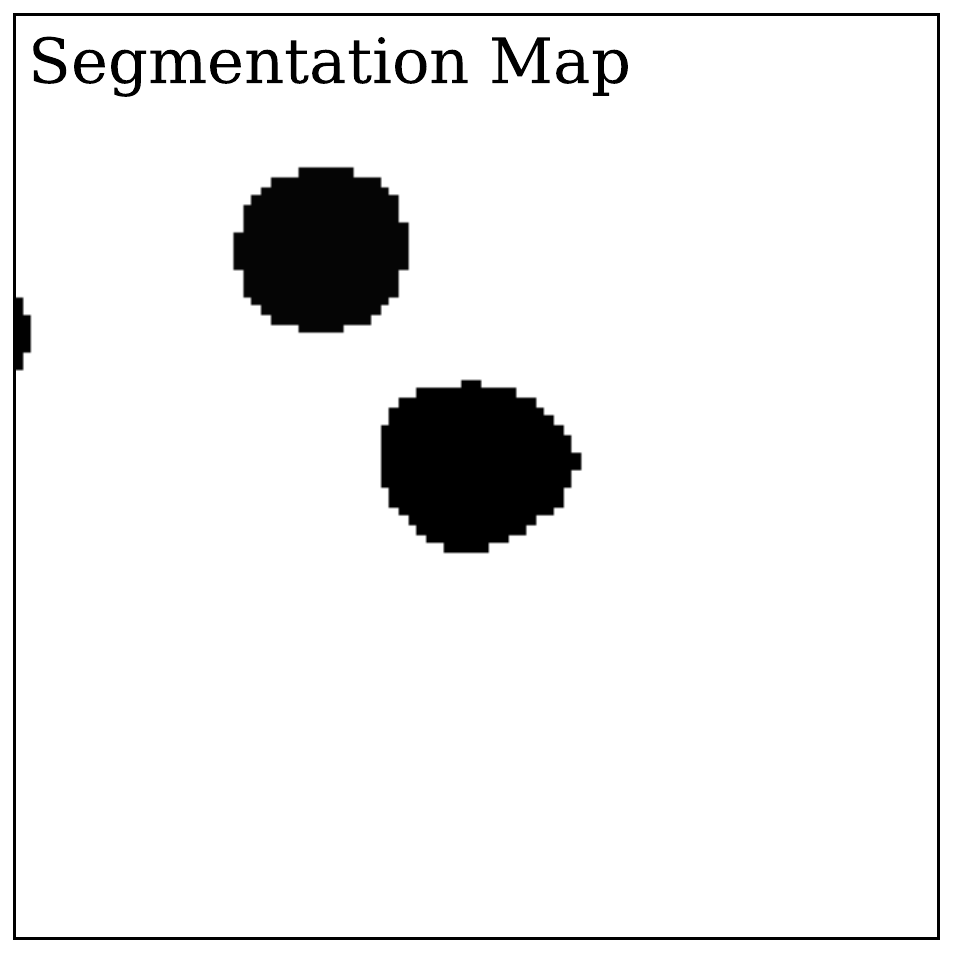}} &
\subfloat{\includegraphics[width = .23\textwidth]{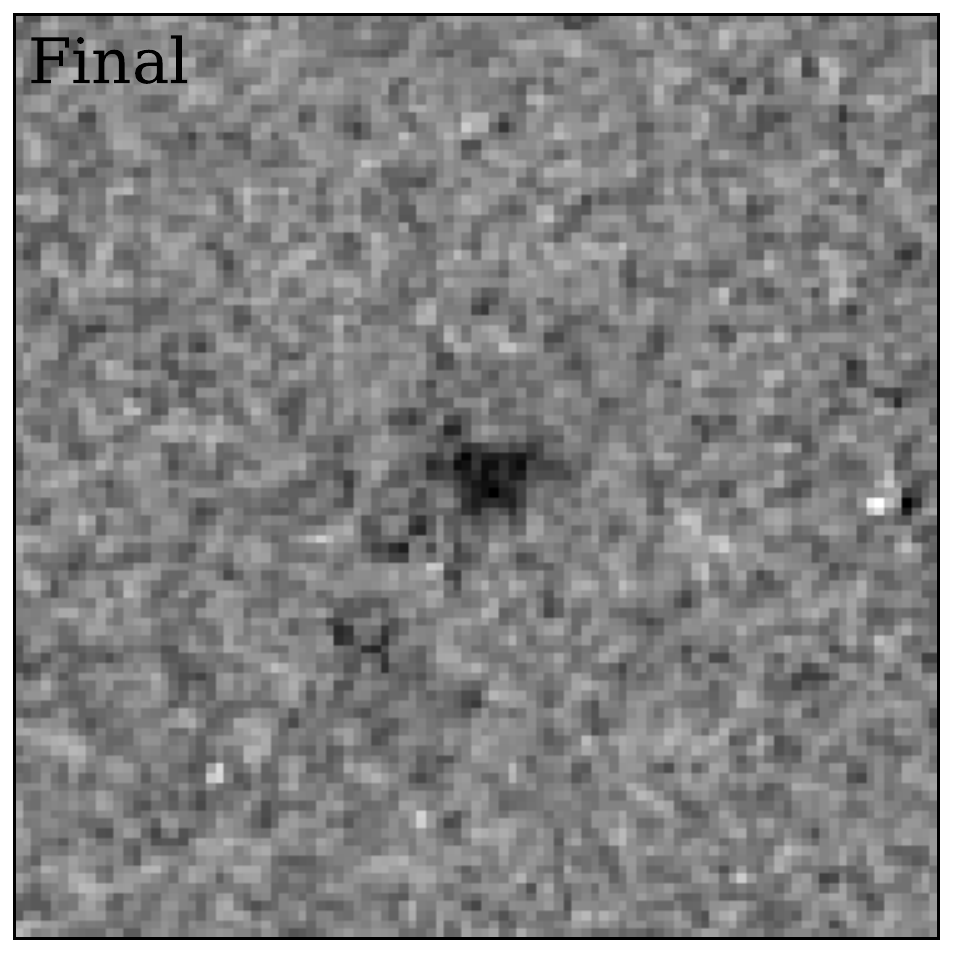}} \\
\end{tabular}
\caption{Examples of our image processing technique for four galaxies at redshifts of 7.0 and 4.5. Each column ($D_{i,j}^{raw}$, $O_{i,j}^{raw}$, $S_{i,j}$, and $O_{i,j}^{analysis}$) corresponds to the parameters of equation \ref{eq:imgproc}. The first column (left) shows the original V$_{606}$ or B$_{435}$ image showing the light below the Lyman-break rest-frame wavelength for the central galaxy's redshift, the second column shows the original H$_{160}$ band image, the third column shows the segmentation map corresponding to the optical rest-frame, while the fourth column (right) shows the result of the image processing whereby all galaxies that appear below the Lyman-break are removed (see equation \ref{eq:imgproc} for details). The field of view is 6'' on a side.}
\label{fig:img_process}
\end{figure*}

\begin{table}
\centering
\caption{The bands used to complete the image processing for each redshift in column 1. Column 2 gives the band corresponding to the UV rest-frame ($O_{i,j}^{raw}$), and Column 3 gives the band corresponding to the Lyman-break where applicable ($D_{i,j}^{raw}$). Column 4 gives the rest frame wavelength probed.}
  \begin{tabular}{cccc}
  \hline 
  \hline
  $z$ & $O_{i,j}^{raw}$ & $D_{i,j}^{raw}$ & $\lambda_{rest}$\\
  \hline
  1 & B$_{435}$ & - & 2175{\AA} \\
  2 & V$_{606}$ & - & 2020{\AA} \\
  3 & I$_{814}$ & - & 2035{\AA} \\
  4 & z$_{850}$ & B$_{435}$ & 1700{\AA} \\
  5 & Y$_{105}$ & B$_{435}$ & 1750{\AA} \\
  6 & J$_{125}$ & V$_{606}$ & 1785{\AA} \\
  \label{tab:imgprocessing}
  \end{tabular}

\end{table}

\subsection{Size Measurement \& Magnitude Determination} \label{sec:method_sizemag}

In this work we use the Petrosian Radius ($R_{\textup{Petr}}(\eta)$) to measure the sizes of our galaxies  \citep[e.g.][]{whitney19}.  The radius we use for sizes, defined by the Petrosian index, is also the radius we use to measure the magnitude of the galaxy.  The Petrosian radius is defined as the radius at which the surface brightness at a given radius is a particular fraction of the surface brightness within that radius \citep[e.g.][]{bershady00, conselice03b}. The radius measured depends on a defined ratio of surface brightness, $\eta(r)$. This ratio is defined as 

\begin{equation}
\eta(r) = \frac{I(r)}{\left\langle I(r) \right\rangle}
\end{equation}

\noindent where $I(r)$ is the surface brightness at radius $r$ and $\left\langle I(r) \right\rangle$ is the mean surface brightness within that radius. By this definition, $\eta(r)$ is 1 at the centre and 0 at large $r$ \citep{kron95}. The Petrosian radius at $\eta = 0.2$ contains at least 99$\%$ of the light within a given galaxy \citep{bershady00}. Throughout this paper, we set the size of a given galaxy to its Petrosian radius at $\eta = 0.2$ and we refer to such size as $R_{\textup{Petr}}$.

The Petrosian radius is determined using the CAS (concentration, asymmetry, and clumpiness) code \citep{conselice03b}. The Petrosian radius differs from the more commonly used half-light radius in that the former is a redshift independent measure of galaxy size. By measuring the sizes of galaxies in this way, we can assume that the measurement would be the same no matter what redshift it was placed whereas by using the half-light radius there is the potential for the size measurement of a particular galaxy to decrease as redshift increases as outer light is lost \citep{buitrago13, whitney19}.  We correct the sizes for point spread function (PSF) effects by simulating a sample of galaxies and applying the WFC3 PSF to images of these galaxies, as described in \cite{whitney19}. All of our galaxies are resolved and as such, the Petrosian radius is never so small that it is dominated by noise rather than galaxy light.

In order to calculate the surface brightness, we first calculate the magnitude, $m$, within an aperture of radius $R_{\textup{Petr}}$ by measuring the flux within this aperture. From this magnitude, we are able to determine the observed surface brightness and the intrinsic surface brightness, as explained below.

\subsection{Surface Brightness Dimming} \label{sec:sbdim}

Below we give a description of our derivation for how the surface brightness of a galaxy changes due to cosmological effects, with a more detailed explanation in Appendix \ref{app:sb_evol}.  There are many papers on this in the past \citep{tolman30, tolman34, giavalisco96, conselice03b} however, we rederive the net correction for the dimming in our own specific situation whereby we are observing the same rest-frame wavelength, but in different observed filters, at different redshifts. Note that all situations in surface brightness dimming are unique and the derivation below is focused on our own situation.

It is well known that the measured surface brightness of a given galaxy is not constant with redshift, and that objects are subject to surface brightness dimming whereby an object at redshift $z_1$ is a factor $f$ dimmer in surface brightness than the same object at redshift $z_2$ where $z_1 > z_2$ \cite{tolman30, tolman34}.  There are 5 factors of $(1+z)$ that need to be taken into account when considering surface brightness dimming; two factors of $(1+z)$ arise from the fact that the source was closer to the observer when the light was emitted. This causes the object to look larger by a factor $(1+z)$ in two dimensions. One factor of $(1+z)$ arises from a change in the rate of photons being received from the source. Another factor is the result of photons shifting to a lower energy as they propagate from the source to the observer. The final factor comes from the change in the unit wavelength bandpass. When considering the integrated flux, the final factor does not apply. This is consistent with the argument made by \cite{tolman30} whereby galaxies dim with redshift by a factor of $(1+z)^4$. Here we consider the flux density measured in units of $f_{\nu}$. Therefore, the redshift dependence is reduced to $(1+z)^3$. 

When considering the spectral flux density in units of $f_{\lambda}$ (erg s$^{-1}$ cm$^{-2}$ \AA$^{-1}$), as used in Space Telescope (ST) magnitudes, we must consider all five factors of $(1+z)$ described above when calculating how much the surface brightness reduces by in terms of a bolometric flux. As such, the intrinsic surface brightess goes as:

\begin{equation}
    \mu_{int} \propto (1+z)^{-5}
\end{equation}

\noindent in units of $erg s^{-1} cm^{-2}$ \AA$^{-1}$, or ST magnitudes, where $\mu_{int}$ is the intrinsic surface brightness.

In the case of the spectral flux density is units of $f_{\nu}$ ($erg s^{-1} cm^{-2} Hz^{-1}$), as used in AB magnitudes, we must consider the first four factors described above (energy, time dilation, angular size). However we must consider the change in unit frequency as opposed to the change in unit wavelength. As frequency is inversely proportional to wavelength, the frequency interval decreases by a factor of $(1+z)$ from emission to detection. From this, we find that 

\begin{equation}
    \mu_{int} \propto (1+z)^{-3}
    \label{eq:absb}
\end{equation}

\noindent in units of $erg s^{-1} cm^{-2}$ Hz$^{-1}$, or AB magnitudes. We use this factor of $(1+z)^{-3}$ throughout the remainder of this work when comparing the measured surface brightness at different redshifts as derived in equation \ref{eq:absb}. 

We use the factor of $(1+z)^4$ when artificially redshifting galaxies, as described below in Section \ref{sec:method_arg}.  Overall, the surface brightness dimming factor used when considering the integrated flux ($erg s^{-1} cm^{-2}$) includes the factors that are a result of energy reduction, time dilation, and change in size. Therefore, objects are subject to a cosmological dimming factor of $(1+z)^{-4}$. This is also the case when considering pixel values of an image. We carry out a detailed derivation in Appendix \ref{app:sb_evol} on a step by step basis so that other situations can be adopted using our methodology.

\subsection{Surface Brightness: Observed and Intrinsic} \label{method:sb}

One of the main goals of this paper is to examine the intrinsic evolution of the surface brightness for galaxies up to $z = 6.5$.  However, this is a derived quantity that must be calculated based on the observed surface brightness, and assuming that cosmological dimming is taking place. As such, we examine in this paper both the intrinsic and the derived intrinsic SB evolution. 

First, we describe the method in which we use to derive the intrinsic surface brightness from the observations. The observed surface brightness is given by: 

\begin{equation}
    \mu_{obs} = m + 2.5\textup{log}_{10}(\pi R_{\textup{Petr}}^2)
    \label{eq:sbobs}
\end{equation}

\noindent where $m$ is the apparent magnitude discussed in Section \ref{sec:method_sizemag} and $R_{\textup{Petr}}$ is the Petrosian radius of the galaxy in which that magnitude is measured. The apparent magnitude is calculated from the measured flux within the measured Petrosian radius $R_{\textup{Petr}}$. As we are measuring fluxes in the AB magnitude system, this surface brightness should be corrected by surface brightness dimming of the form:

\begin{equation}
    \mu_{int} = m + 2.5\textup{log}_{10}(\pi R_{\textup{Petr}}^2) - 2.5\textup{log}_{10}((1+z)^3).
    \label{eq:sbint}
\end{equation}

\noindent We use eq. \ref{eq:sbint} to calculate the intrinsic SB for the systems we detect, within the observed SB given by eq. \ref{eq:sbobs}. Throughout the rest of this paper we use these SB values to understand the physical evolution and selection effects present within deep surveys such as CANDELS. As explained in Section \ref{sec:sbdim}, we use a factor of $(1+z)^4$ to determine the amount of cosmological dimming experienced by a galaxy when being artificially redshifted from a redshift $z_{low}$ to a higher redshift $z_{high}$.  

\subsection{Artificially Redshifted Galaxies} \label{sec:method_arg}

One of the tools we use to determine whether galaxies observed at low redshift ($z_{low}$) would be detectable at higher redshift ($z_{high}$), is by artificially redshifting galaxies in our sample from low redshift to high redshift.  We start by examining a representative simulation by taking galaxies from $z_{low} = 2$ and simulating how they would appear within the CANDELS data at $z_{high} = 6$. This procedure is done by multiplying the V$_{606}$ image by the factor $\left(\frac{1+z_{lo}}{1+z_{hi}}\right)^4$ and inserting this into the background of the J$_{125}$ mosaic image whose noise is reduced by a factor defined as,

\begin{equation}
    f = \frac{1}{1-\left(\frac{1+z_{lo}}{1+z_{hi}}\right)^4},
\end{equation}

\noindent to ensure we are not overestimating the background. We also include luminosity evolution in the form of the difference between the characteristic magnitudes ($\Delta M^*$) of the UV luminosity functions for each redshift. The values for $1 \leq z \leq 3$ are taken from \cite{arnouts05} and the values for $4 \leq z \leq 7$ are taken from \cite{bouwens15}. The redshifting process is therefore described by the equation

\begin{equation}
    R_{i, j} = UV_{i, j} \cdot \left(\frac{1+z_{lo}}{1+z_{hi}}\right)^4 \cdot \Delta M^* + \frac{N_{i, j}}{f}
\end{equation}

\noindent where $R_{i, j}$ is the redshifted image, $UV_{i, j}$ is the rest frame UV image at $z = 2$, and $N_{i, j}$ is the background noise. We then measure these galaxies using the CAS code \citep{conselice03b} in the same way would do on actual galaxies. We use this to measure the size (in this case, the Petrosian radius at $\eta = 0.2$, $R_{\textup{Petr}}$ and flux of each galaxy from which we can determine the measured surface brightness.

\begin{figure*}
\centering
\includegraphics[width = \textwidth]{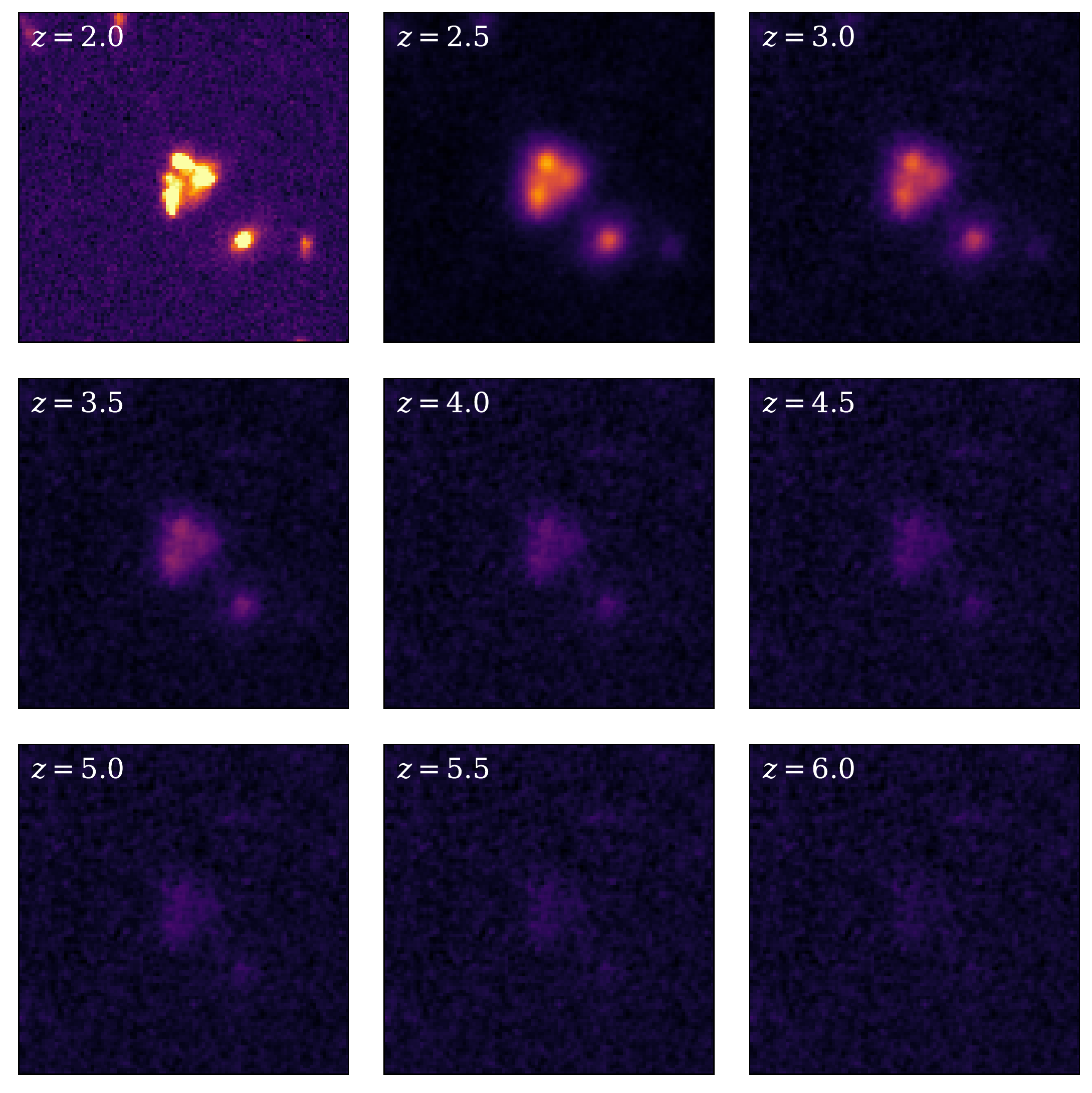}
    \caption{Example of the redshifting process beginning with a $z = 2$ galaxy (top left) and ending with the same galaxy redshifted to $z = 6$ (bottom right). The images in between correspond to the original galaxy redshifted in $\delta z = 0.5$ intervals. The image at each redshift interval has been evolved with the characteristic magnitude of that redshift \citep{arnouts05, bouwens15}.}
    \label{fig:redshiftingex}
\end{figure*}

We calculate a surface brightness completeness limit by examining the whole sample from which the two sub-samples are selected and determining at which magnitude the surface brightness function declines. This magnitude is then used to calculate the equivalent intrinsic surface brightness at each redshift.  

\section{Results} \label{sec:results}

In this Section we present the results we find when measuring the surface brightness of the galaxies using the two separate samples described in Section \ref{sec:data}; a mass-selected sample where the galaxies lie within a mass range of 10$^{10}$M$_{\odot}$$\leq$M$_*$$\leq$10$^{11}$M$_{\odot}$ and a number density-selected samples where each redshift bin contains galaxies such that a number density selection is used at the constant value of 1$\times$10$^{-4}$ Mpc$^{-3}$ with a mass range of $10^{9.5} M_{\odot} < M_* < 10^{11.8}M_{\odot}$. This particular method of selecting galaxies allows us to trace how the most massive galaxies have evolved in SB over time.

\subsection{Observed Surface Brightness as Function of Redshift}

Throughout this work we measure all sizes and fluxes in the UV rest-frame between 1785\AA, and 2175\AA. Surface brightness is variant with wavelength, hence the need for a consistent rest-frame wavelength across redshifts. The UV rest-frame is also useful for tracing the star formation within galaxies. One downside of using the UV is that we are not able to trace the underlying older stellar mass formed within these galaxies. We are only tracing the young stars and the effects of dust on this light. 

\begin{figure}[!ht]
\includegraphics[width=0.475\textwidth]{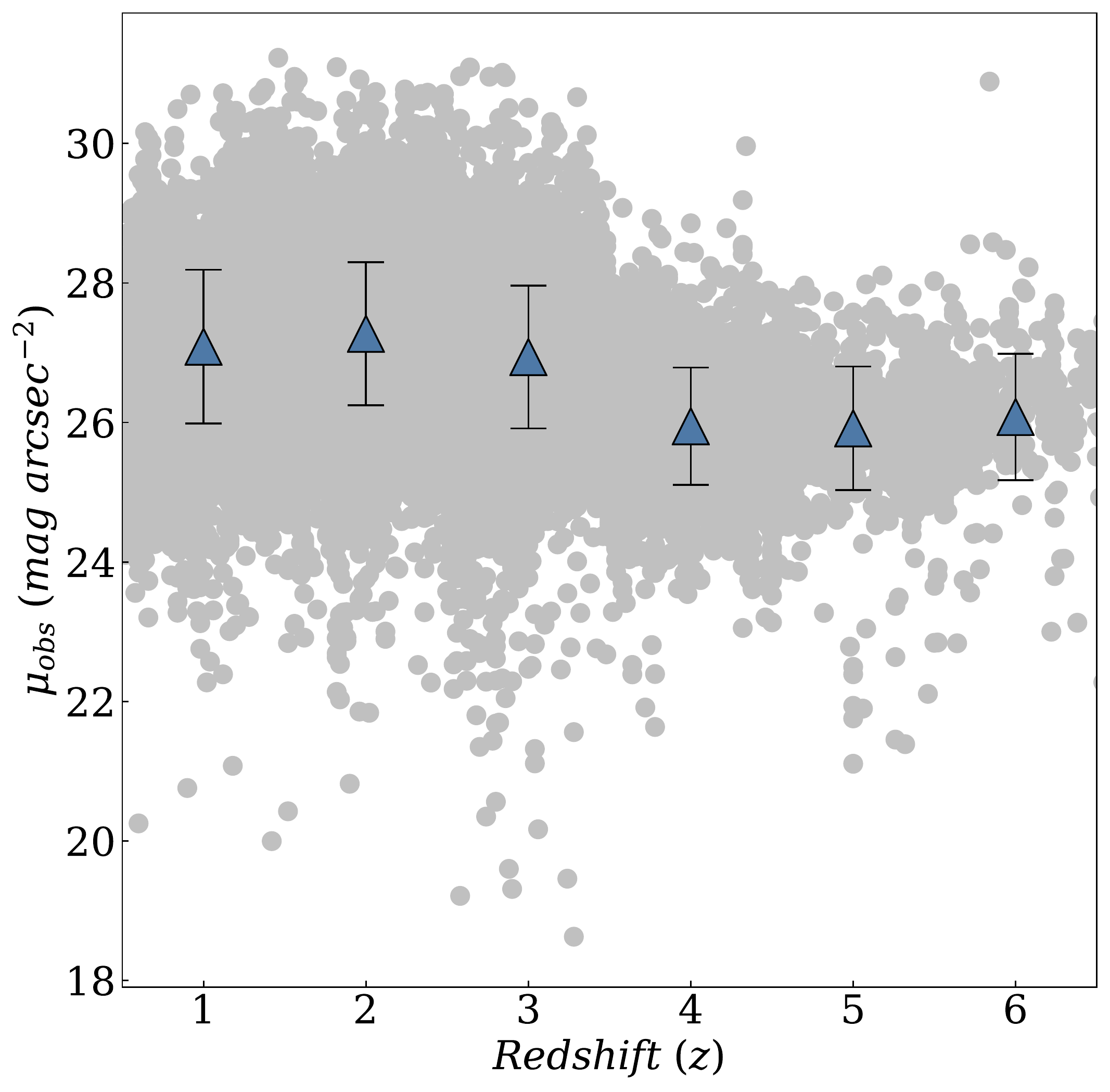}
\caption{Observed surface brightness for the full sample from which the mass-selected and number density-selected samples are taken from. The average observed surface brightness for each redshift bin are shown as blue triangles with errors equal to 1$\sigma$. This is purely an observational result showing the range of surface brightness which we probe with our filters and exposures.}
\label{fig:fullobssb}
\end{figure}

Before applying mass and number density cuts to the data, we examine the observed surface brightness distribution of the full sample of $\sim$34,000 galaxies obtained from the CANDELS GOODS-North and GOODS-South fields. These galaxies lie in the redshift range $0.5 < z < 6.5$ and have stellar masses in the range $10^{6}M_{\odot} < M_* < 10^{12}M_{\odot}$. The observed surface brightness for each of these galaxies can be seen in Figure \ref{fig:fullobssb} before any cuts are performed. The average observed surface brightness of each redshift bin is given as a blue triangle, with an error of 1$\sigma$. There is a jump in $\mu_{obs}$ between redshifts $z = 3$ and $z = 4$ due to a difference in limiting magnitude of the filters, whereby F814W probing $z \sim 3$ is more sensitive than any redder filter with WFC3.

\begin{figure*}
\centering
\begin{tabular}{cc}
\subfloat{\includegraphics[width = .49\textwidth]{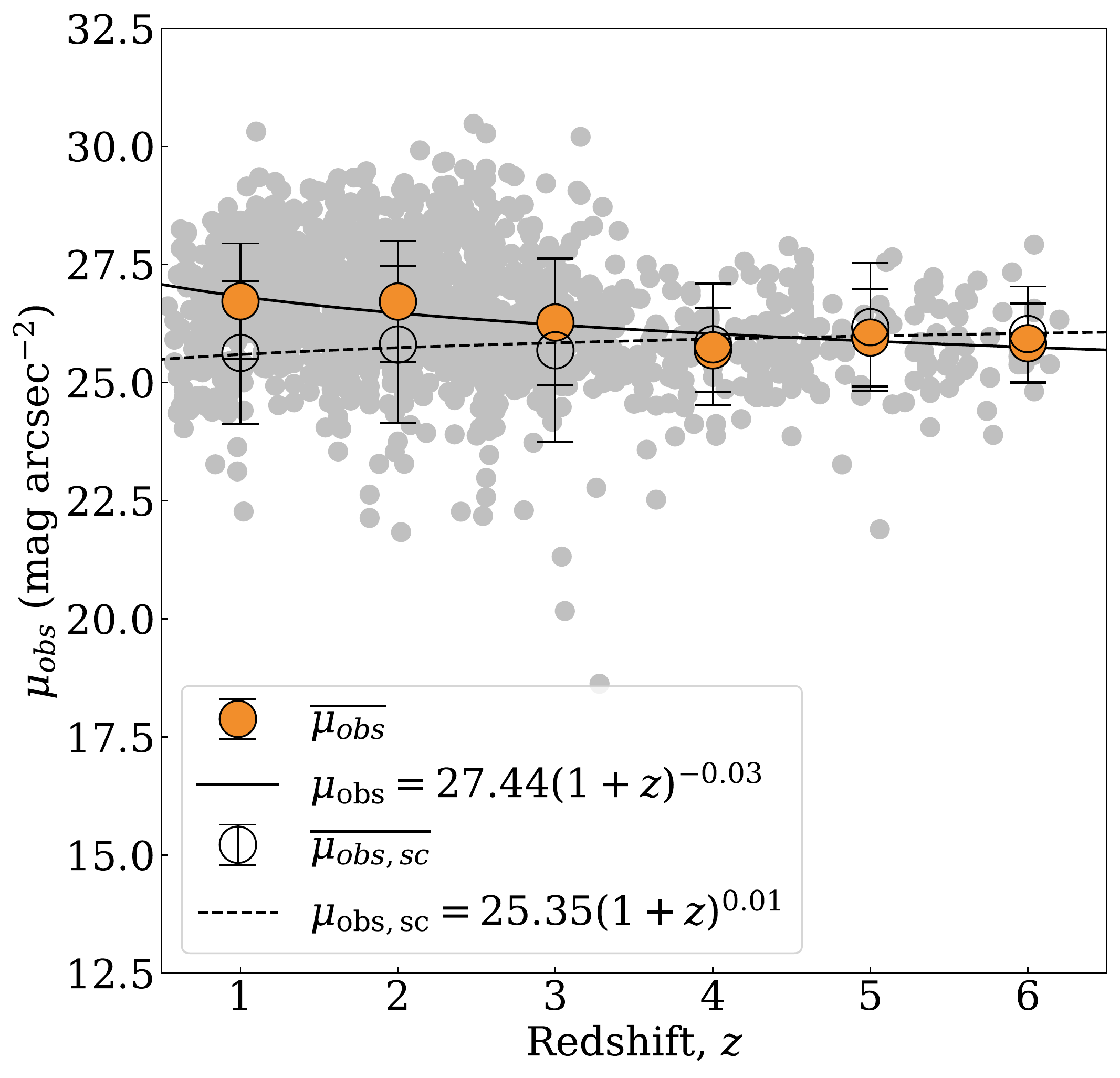}} &
\subfloat{\includegraphics[width = .49\textwidth]{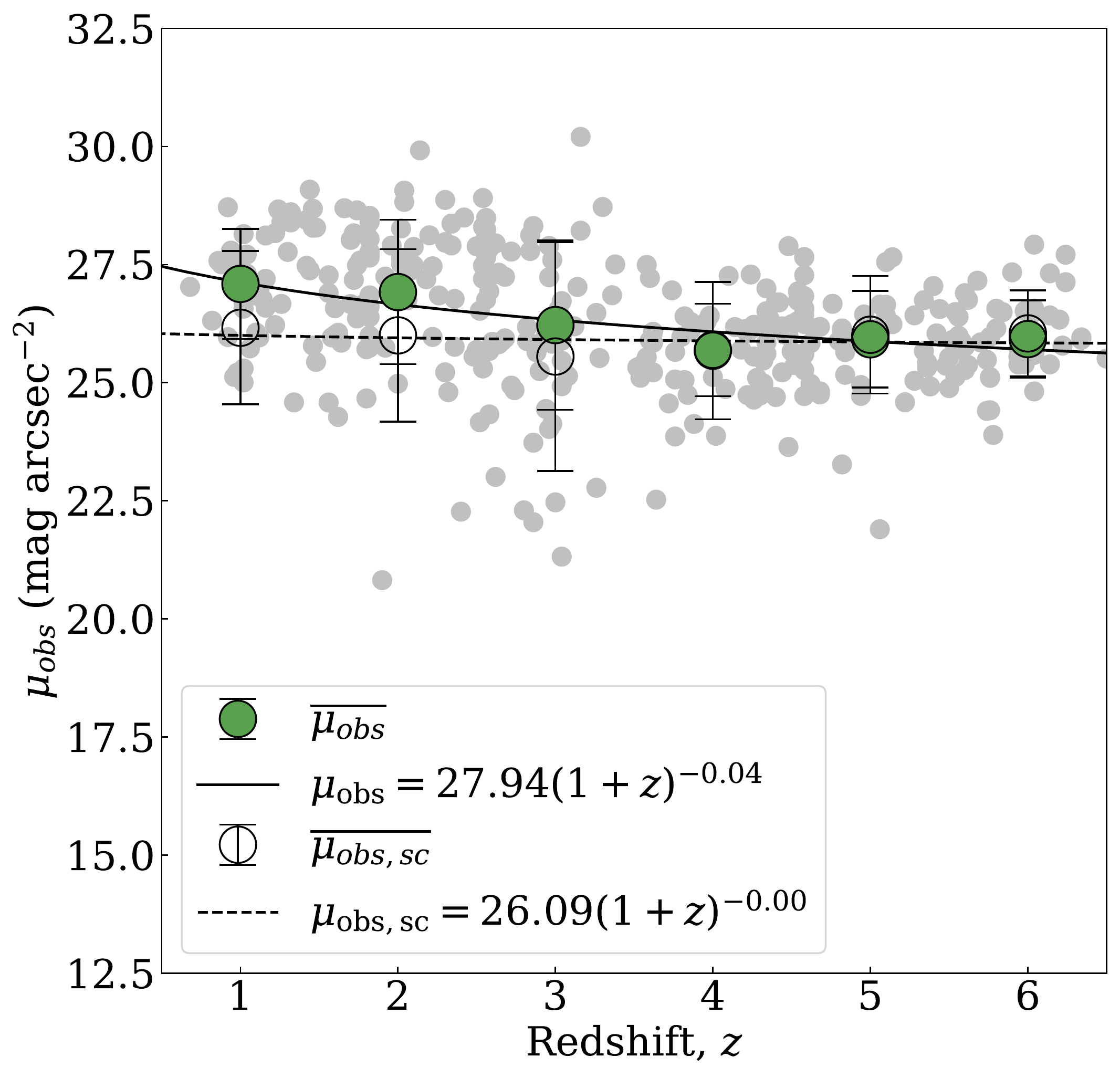}}
\end{tabular}
\caption{Left: evolution of the average observed surface brightness for the mass-selected sample. The mean surface brightness is shown for each redshift as orange circles. A power law is fit to the mean values and we find $\mu_{\textup{obs}} = 27.44 \pm 0.34 (1+z)^{-0.03 \pm 0.01}$. This fit is shown as a solid line. The surface brightness is corrected for size by setting the sizes of all galaxies to the mean $z = 6$ galaxy size . These points are shown as empty circles. The fit to the points is shown as a dashed line and is found to be $\mu_{\textup{obs}} = 25.35 \pm 0.20 (1+z)^{0.01 \pm 0.01}$. The error bars shown are 1$\sigma$ from the mean. Right: Evolution of the average observed surface brightness for the number density-selected sample. The mean surface brightness for each redshift is indicated by the green circles. The error bars shown are 1$\sigma$ from the mean and a fit of the form $(1+z)^{-\beta}$ (solid line) have been fit to the data. The fit is found to be $\mu_{\textup{obs}} = 27.94 \pm 0.39 (1+z)^{-0.04 \pm 0.01}$. The observed surface brightness is corrected for size, as shown by the empty circles. The power law fit of these values yields $\mu_{\textup{obs}} = 26.09 \pm 0.37 (1+z)^{-0.00 \pm 0.01}$ and is shown as a dashed line.}
\label{fig:obssb}
\end{figure*}

The evolution of the observed surface brightness ($\mu_{obs}$) of the mass-selected sample can be seen in the left panel of Figure \ref{fig:obssb} with the mean surface brightness indicated for each redshift as orange circles. The same plot for the number density-selected sample can be seen in the right panel of Figure \ref{fig:obssb} with the mean surface brightness shown by green circles. Both forms of the evolution are fit with a power law of the form:

\begin{equation}
    \mu = \alpha(1+z)^{-\beta}.
    \label{eq:powerlaw}
\end{equation}

\noindent The parameters determined for the fits are shown in Table \ref{tab:sbfits}. Both samples show a relatively flat evolution, increasing by 0.9 $\pm$ 1.5 mag arcsec$^{-2}$ and 1.2 $\pm$ 1.4 mag arcsec$^{-2}$ respectively. We correct the observed surface brightness for size evolution by setting the size of each galaxy to be the mean size of the $z = 6$ galaxies such that eq. \ref{eq:sbobs} becomes 

\begin{equation}
    \mu_{obs} = m + 2.5\textup{log}_{10}(\pi R_{\textup{z=6}}^2).
    \label{eq:sbobs_z6}
\end{equation}

\noindent We plot these size corrected values as empty circles and fit these points with the power law given in eq. \ref{eq:powerlaw} and this is shown as a dashed line on both the left and right panels of Figures \ref{fig:obssb}. We see that the size corrected values yield a shallower evolution whereby the difference between the surface brightness at $z=1$ and $z = 6$ is now 0.4 $\pm$ 1.8 mag arcsec$^{-2}$ and 0.1 $\pm$ 1.9 mag arcsec$^{-2}$ for the mass-selected and number density-selected samples respectively. 

When using more the more direct physical value of surface brightness in the form of measured flux per unit area in real units rather than logged, we find that the mass-selected sample evolves as $(1+z)^{0.25 \pm 0.71}$ and the number density-selected sample evolves as $(1+z)^{0.07 \pm 0.73}$. This method also suggests that the flux per unit area decreases with time however this evolution is much steeper than the evolution in logged units, particularly in the case of the mass-selected sample.

\subsection{Intrinsic Surface Brightness as a Function of Redshift}

\begin{figure*}
\centering
\begin{tabular}{cc}
\subfloat{\includegraphics[width = .49\textwidth]{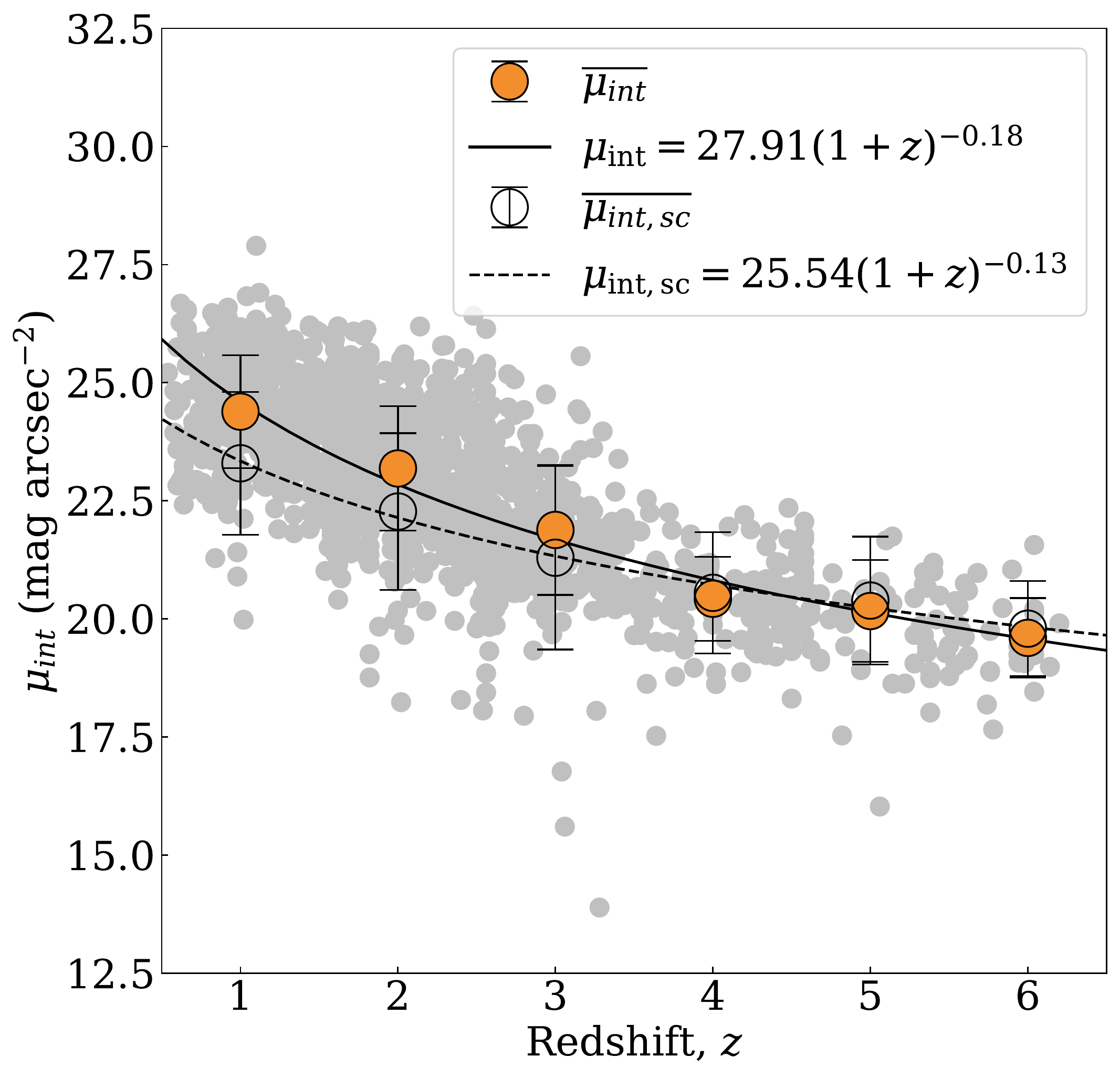}} &
\subfloat{\includegraphics[width = .49\textwidth]{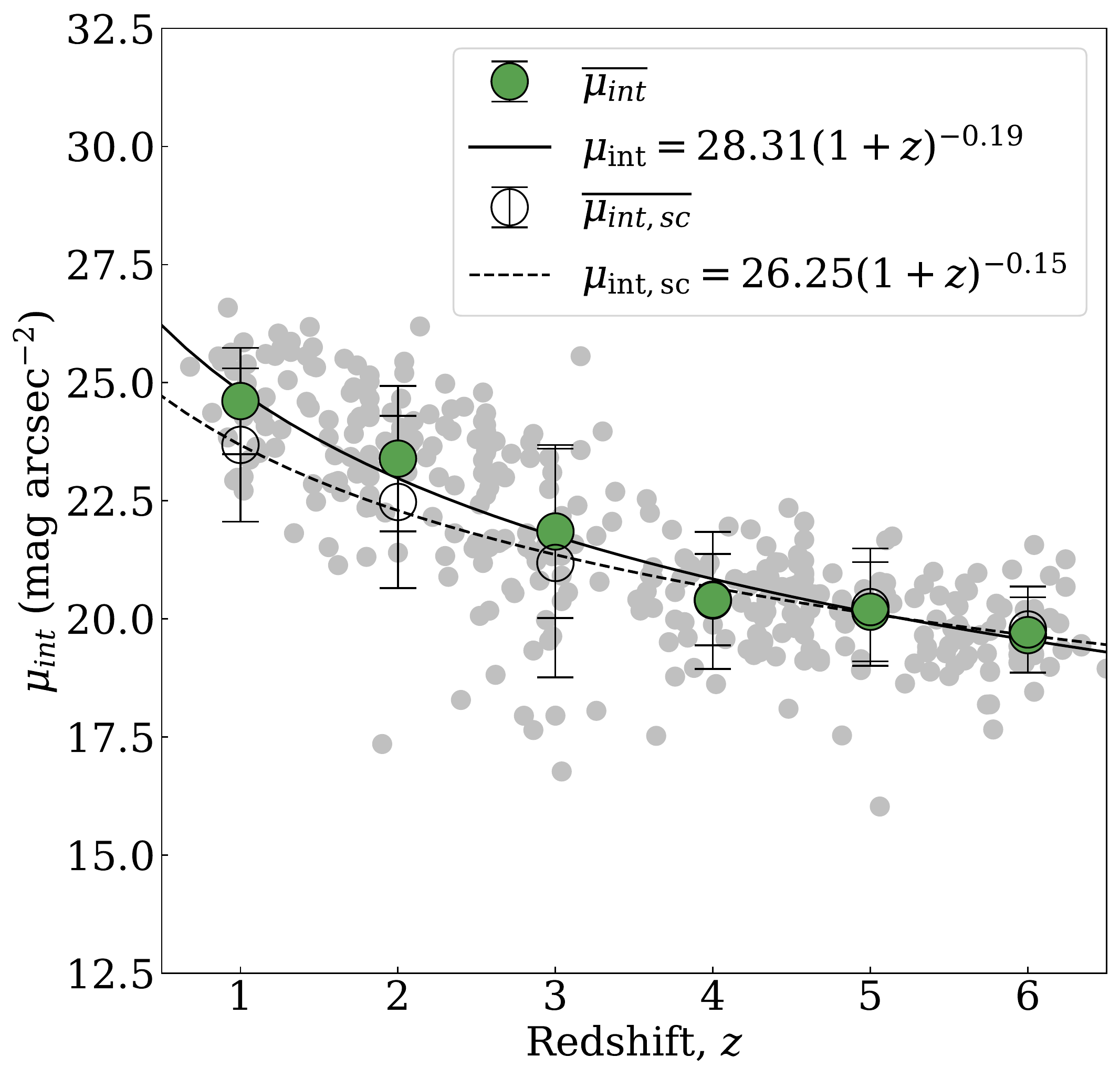}} 
\end{tabular}
\caption{Left: evolution of the intrinsic surface brightness for the mass-selected sample with the mean surface brightness for each redshift bin shown by the orange circles. We fit a power law of the form $(1+z)^{-\beta}$ (solid line) and find $\mu_{\textup{int}} = 27.91 \pm 0.51 (1+z)^{-0.18 \pm 0.01}$. Also shown are the mean size corrected values (empty circles) and a power law fit to these points. This fit is found to be $\mu_{\textup{int}} = 25.54 \pm 0.22 (1+z)^{-0.13 \pm 0.01}$. The error bars shown are 1$\sigma$ from the mean. Right: evolution of the intrinsic surface brightness for the number density-selected sample with the mean surface brightness for each redshift bin shown by the green circles. We fit a power law (shown by the solid line) and find $\mu_{\textup{int}} = 28.31 \pm 0.57 (1+z)^{-0.19 \pm 0.01}$. The mean size corrected surface brightness is also shown as empty circles, along with the fit to these points. The fit is found to be $\mu_{\textup{int}} = 26.25 \pm 0.34 (1+z)^{-0.15 \pm 0.01}$. The error bars shown are 1$\sigma$ from the mean.}
\label{fig:intsb}
\end{figure*}

The intrinsic surface brightness ($\mu_{int}$) varies with redshift such that galaxies at a low redshift are $\sim$5 mag dimmer than those at the highest redshift. The evolution of $\mu_{int}$ can be seen in the left panel of Figure \ref{fig:intsb} for the mass-selected sample and in the right panel of Figure \ref{fig:intsb} for the number density-selected sample. The mean intrinsic surface brightness is indicated by the orange and green points respectively. 

We fit a power law to these mean values and these fits are indicated by a solid line in both cases. The parameters of the power law fits for both figures are shown in Table \ref{tab:sbfits}.  Both samples show a similar evolution, with the mass-selected sample evolving as $(1+z)^{-0.18 \pm 0.01}$ and the number density-selected sample evolving as $(1+z)^{-0.19 \pm 0.01}$. No matter the selected method, the intrinsic surface brightness changes by several mag arcsec$^{-2}$ with the mass-selected sample changing by 4.8 $\pm$ 1.5 mag arcsec$^{-2}$ and the number density-selected sample changing by 5.0 $\pm$ 1.4 mag arcsec$^{-2}$. 

As for the observed surface brightness, we correct for size evolution for both samples by setting the size of each galaxy to the mean size of the $z = 6$ bin. Eq. \ref{eq:sbint} therefore becomes

\begin{equation}
    \mu_{int} = m + 2.5\textup{log}_{10}(\pi R_{\textup{z=6}}^2) - 2.5\textup{log}_{10}((1+z)^3).
    \label{eq:sbint_z6}
\end{equation}

\noindent This is shown in the left (mass-selected) and right (number density-selected) panels of  Figure \ref{fig:intsb} by the empty circles and a fit of the form $(1+z)^{-\beta}$ is shown as a dashed line for both samples. This size correction causes the evolution to flatten for both samples with $\beta$ changing from 0.18 $\pm$ 0.01 to 0.13 $\pm$ 0.01 and 0.19 $\pm$ 0.01 to 0.15 $\pm$ 0.01 for the mass-selected and number density-selected samples respectively. This size correction also causes the difference in surface brightness between $z = 1$ and $z = 6$ to change to 3.5 $\pm$ 1.8 mag arcsec$^{-2}$ and 3.9 $\pm$ 1.9 mag arcsec$^{-2}$ for the two samples. This means that size can only account for about a magnitude of the evolution of galaxy surface brightness, with 3-4 mag arcsec$^{-2}$ unaccountable for the fact that galaxies are growing in size within their Petrosian radius as they evolve from high to low redshift \citep[e.g][]{whitney19}. By examining the evolution of the size corrected surface brightness, a quantity that is linearly proportional to the luminosity, we are effectively examining the evolution of the absolute magnitude at a fixed size.

\begin{table}\centering
\caption{The fits determined for both the mass-selected (M) and number density-selected (ND) samples as given by equation \ref{eq:powerlaw}.}
  \begin{tabular}{cccc}
  \hline 
  \hline
   & Sample & $\alpha$ & $\beta$ \\
  \hline
  \multirow{2}{*}{$\mu_{\textup{observed}}$} & M & 27.44 $\pm$ 0.34 & -0.03 $\pm$ 0.01 \\
                                             & ND & 27.94 $\pm$ 0.39 & -0.04 $\pm$ 0.01 \\
  \multirow{2}{*}{$\mu_{\textup{intrinsic}}$} & M & 27.91 $\pm$ 0.51 & -0.18 $\pm$ 0.01 \\
                                             & ND & 28.31 $\pm$ 0.57 & -0.19 $\pm$ 0.01 \\
  \label{tab:sbfits}
  \end{tabular}

\end{table}

\begin{figure}[!ht]
\includegraphics[width=0.475\textwidth]{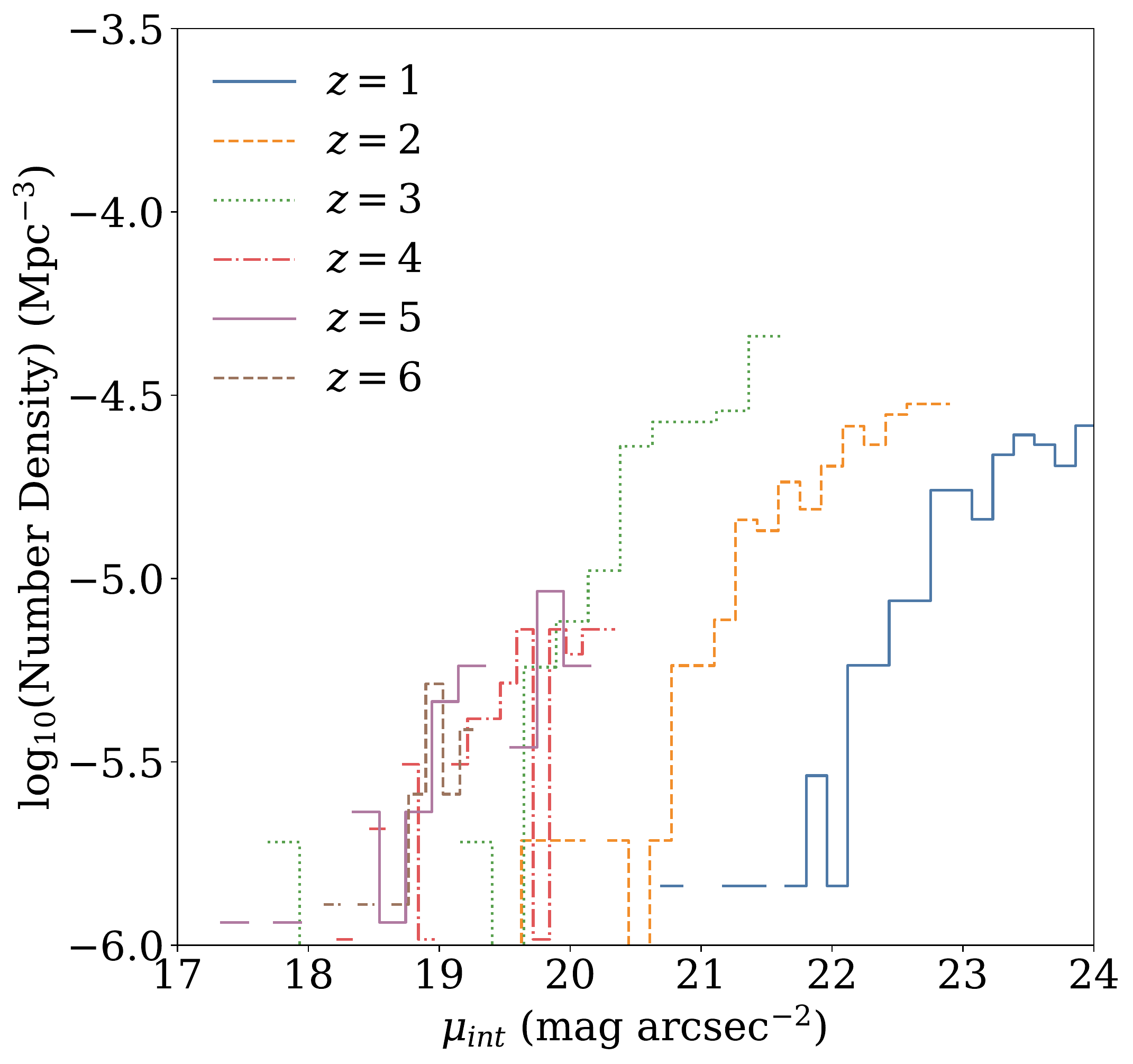}
\caption{Intrinsic surface brightness distribution for each redshift. The different coloured lines show the different redshifts for this distribution of intrinsic surface brightness. These are corrected by a factor of (1+z)$^3$ from the observed distributions to account for the surface brightness dimming with redshift.
}
\label{fig:intsb_fit}
\end{figure}

We also show the distribution of the surface brightness for the full mass-selected sample in Figure \ref{fig:intsb_fit}. This demonstrates a systematic evolution at all galaxy masses in surface brightness, such that, on average, galaxies at higher redshifts exhibit a higher intrinsic surface brightness, which then declines at lower redshifts. One caveat to this observation is that this is not necessarily a complete sample, as we would naturally be missing galaxies lower than the surface brightness completeness limit.  At high redshift this limit is quite high -- 20 mag arcsec$^{-2}$, which appears to be the limit in which we can still detect intrinsically faint galaxies at $z > 3$.  What remains to be known or determined is if there are indeed galaxies at these redshifts which have an intrinsic surface brightness which is lower than this value and therefore unobservable with our current deep imaging.  We investigate this question later in the paper.

\subsection{Artificially Redshifting Galaxies} \label{sec:arg}

We examine the effects of redshift by simulations described in \S \ref{sec:method_arg}.  We carry out these simulations by artificially redshifting galaxies at low redshifts at $z = 2$ and imaging then to how they would appear in our data at $z = 6$. This is done primarily using the techniques described in \S \ref{sec:method_arg} whereby the galaxy is reduced in surface brightness by (1+z)$^{4}$, and then reimaged into our data.

When we carry out this process we find that there are few galaxies obviously visible in the resulting images. This implies that without some type of luminosity evolution we would only be able to detect a few galaxies with properties similar to the ones we can detect at $z = 2$ at higher redshifts. However, in addition to our visual determination of whether a galaxy has been detected, we measure the signal-to-noise ratio ($S/N$) of each of the objects after they are simulated. We measure the flux within $R_{\textup{Petr}}(\eta = 0.2)$. The average $S/N$ for the redshifted sample is $\sim$ 3, which is barely detectable at best, and which is a significant decrease when compared to the original sample of galaxies that have an average $S/N$ of $\sim$ 100. We find that 16\% of the artificially redshifted galaxies are detectable (defined as having a $S/N$ $>$ 5) compared to 94\% of the original sample of $z = 2$ galaxies. 

We measure a difference of 3.5 $\pm$ 1.6 mag arcsec$^{-2}$ in intrinsic surface brightness between $z = 6$ and $z = 2$ for the mass-selected sample and so we apply this difference to the $z = 6$ simulated galaxies in the form of an increase of brightness by a factor of $\sim 24$ using the relation:

\begin{equation}
    \frac{L_{z=6}}{L_{z=2}} = 10^{0.4\cdot\Delta M}. 
\end{equation}

\noindent As a result, the $S/N$ of the artificially redshifted galaxies increases to about $\sim 50$ after this luminosity evolution is included. We also measure a mean observed surface brightness of 26.3 mag arcsec$^{-2}$ and a mean intrinsic surface brightness of 19.9 mag arcsec$^{-2}$. Both results are $\sim$ 0.3 magnitudes dimmer than the values we see for the actual $z = 6$ sample of galaxies. This implies that most of these galaxies would be observable at $z = 6$ just using the amount of decrease in the observed surface brightness and correcting for it.

However, we know from various studies of the luminosity function in the UV that the intrinsic average brightness of galaxies on average changes as we go to higher redshifts, with galaxies becoming brighter \citep[e.g.][]{arnouts05, bouwens06, mclure13, duncan14, bouwens15, bhatawdekar19}. This observed brightening of the average galaxy population is however much less than the amount we observe in terms of the SB evolution.   

To see the effect of this on the detectability of our sample of galaxies after being simulated from $z = 2$ to $z = 6$ we increase the brightness of our images by the observed amount from LF evolution.  We take the $z = 2$ characteristic magnitude, M$^*_{UV}$, to be $-20.33\pm0.50$ \citep{arnouts05} and set this value to be the zero-point. We take the $z = 6$ characteristic magnitude to be $-20.94\pm0.20$ \citep{bouwens15} and calculate the change in magnitude between the two redshifts to be 0.61 $\pm$ 0.54 mag. We are then able to calculate the factor by which the brightness changes ($\frac{L_{z=6}}{L_{z=2}}$) from the lowest redshift to the highest redshift. We determine this value to be 1.75 $\pm$ 0.87, which is much less than the factor of 24 observed, as discussed above. We calculate this factor for each redshift interval and multiply the redshifted images by this factor to simulate the evolution of the luminosity function. 

\begin{figure}[!ht]
\includegraphics[width=0.475\textwidth]{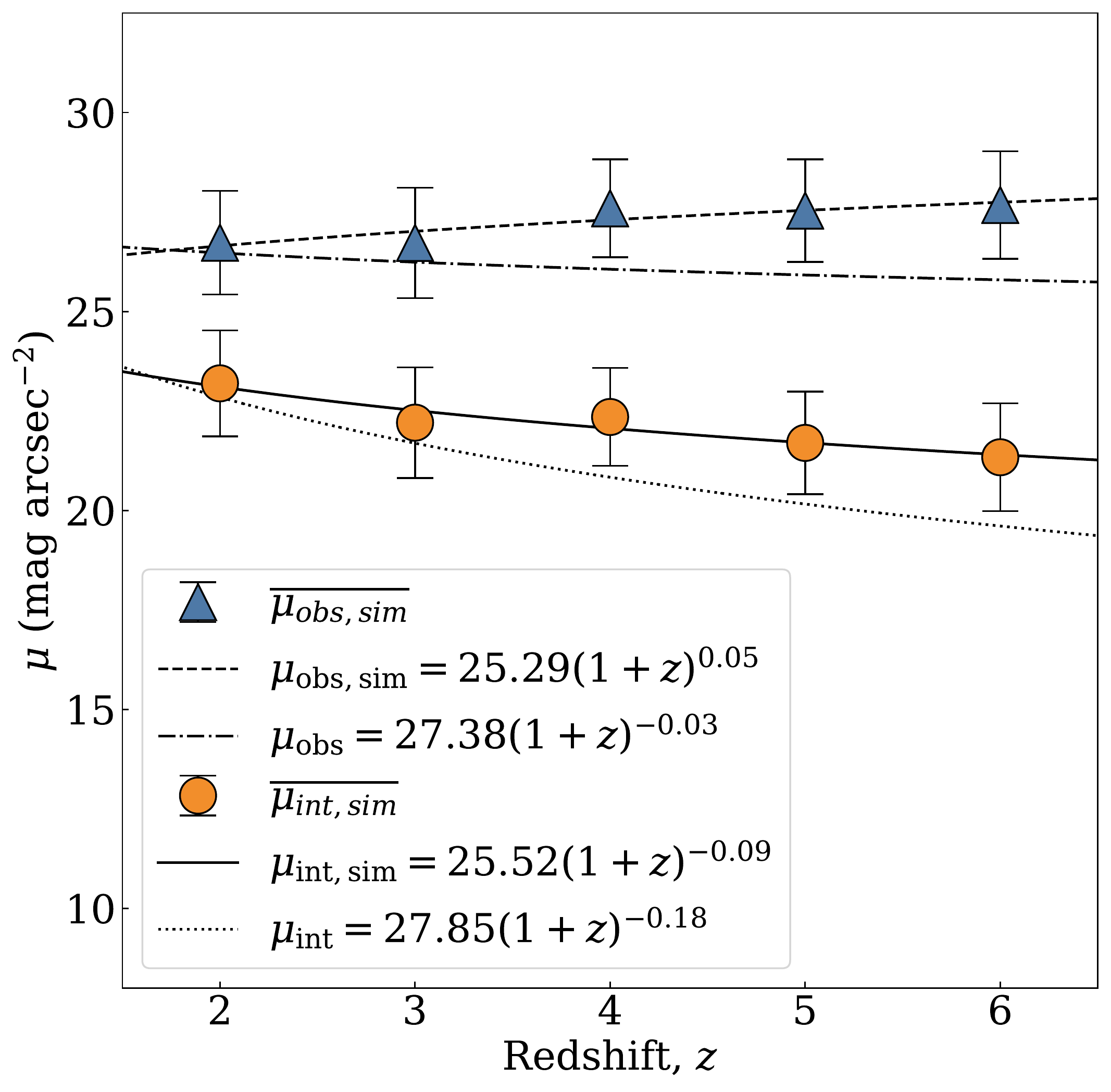}
\caption{Evolution of the observed (blue triangles) and intrinsic (orange circles) surface brightness of our simulated galaxies that have been evolved with the luminosity function increase in brightness from $z = 2$ to $z = 6$. We fit a power law to both surface brightness and find $\mu_{\textup{obs}} = 25.29 \pm 0.30 (1+z)^{0.05 \pm 0.01}$ (shown as a dashed line) for the observed surface brightness and find $\mu_{\textup{int}} = 25.52 \pm 0.67 (1+z)^{-0.09 \pm 0.02}$ (shown as a solid line) for the intrinsic surface brightness. Also shown are the fits to the real surface brightness evolution for both the intrinsic (dotted line) and observed (dot-dash line).}
\label{fig:simsb}
\end{figure}

We show the evolution of both the observed and intrinsic surface brightness of the artificially redshifted sample of galaxies that have been evolved with the luminosity function in Figure \ref{fig:simsb}. Unlike the actual observed surface brightness evolution, the measured SB for the simulated sample increases by 1.0 $\pm$ 1.9 mag arcsec$^{-2}$ from $z = 6$ to $z = 2$, and we find that the evolution goes as $(1+z)^{0.05 \pm 0.01}$ so therefore these simulated galaxies appear to get brighter with redshift. This is due to the method used when artificially redshifting the galaxies; the overall factor the images are multiplied by decreases as redshift increases leading to a smaller measured flux and therefore a dimmer surface brightness. The real mass-selected sample on the other hand decreases by 0.7 $\pm$ 1.6 mag arcsec$^{-2}$ over the same redshift range. The intrinsic surface brightness however follows the trend seen in the real sample but the evolution is not as steep; the evolution goes as $(1+z)^{-0.09 \pm 0.02}$ and the surface brightness decreases by 1.9 $\pm$ 1.9 mag arcsec$^{-2}$ over the redshift range $z = 6$ to $z = 2$ whereas the real sample decreases by 3.5 $\pm$ 1.6 mag arcsec$^{-2}$ over the same redshift range.

Based on this, there is approximately 1.6 magnitudes of intrinsic surface brightness evolution unaccounted for in the simulated images. We find from this that 84\% of the z = 2 galaxies would not be detected at z = 6 if evolved with the luminosity function (where detected galaxies are determined as being those with a S/N $>$ 5), with only the highest surface brightness ones being detectable. This implies that there is a either a significant amount of galaxies being missed in deep HST imaging which exist high redshifts, or that some galaxies have evolved significantly more than others. We will further discuss this in the discussion section of this paper. 

We also alternate how much evolution we add to determine how much brighter the galaxies at $z = 2$ would need to be in order to be considered significant detections. We find that a factor of 6 is required for a mean $S/N$ value of 10. This factor of 6 in brightness equates to a change in magnitude of $\sim 2$ which is much larger than the change of 0.61 in magnitude seen for the characteristic magnitudes between $z = 2$ and $z = 6$. Thus it cannot be the case that surface brightness evolution is determined simply by an evolving luminosity and size within the Petrosian radius. 

The limiting magnitudes for an extended source for the J$_{125}$ filter of GOODS-South and GOODS-North fields are 27.9 mag and 28.0 mag respectively. These values are 5 $\times$ the photometric error within a 0.2 arcsec$^{2}$ aperture \citep{grogin11}. This equates to limiting observed surface brightness to 26.2 mag arcsec$^{-2}$ and 26.3 mag arcsec$^{-2}$ within a 0.2 arcsec$^2$ area, respectively.  However, this is not a proper limit due to various factors.  We thus empirically determine the SB completeness limit for our data by examining at which magnitude the SB function declines. We find that 90\% of the simulated galaxies that have been evolved with the luminosity function correction have an observed surface brightness that is lower than the limiting surface brightness. Therefore, if these galaxies were real, we would not be able to detect the vast majority of these galaxies using the J$_{125}$ filter of WFC3. 

In the next section of this paper we investigate the relationship of surface brightness to other galaxy properties.  One reason we do this is to try to make sense of this evolution in observed surface brightness and what it might imply regarding the galaxy population at high redshift.

\subsection{Correlation With Other Parameters}

This first part of this paper is about the use of surface brightness measures as a way to detect galaxies.  We discuss how it is likely that many galaxies are likely missing and whether the evolution of luminosity is consistent with the intrinsic SB evolution.  In this section, we explore the relationship between the intrinsic surface brightness and other galaxy parameters such as star formation rate. This is the second part of this paper whereby we examine how the surface brightness reveals information about the physical state of the galaxies and how they are evolving.

\begin{figure*}
\centering
\includegraphics[width = \textwidth]{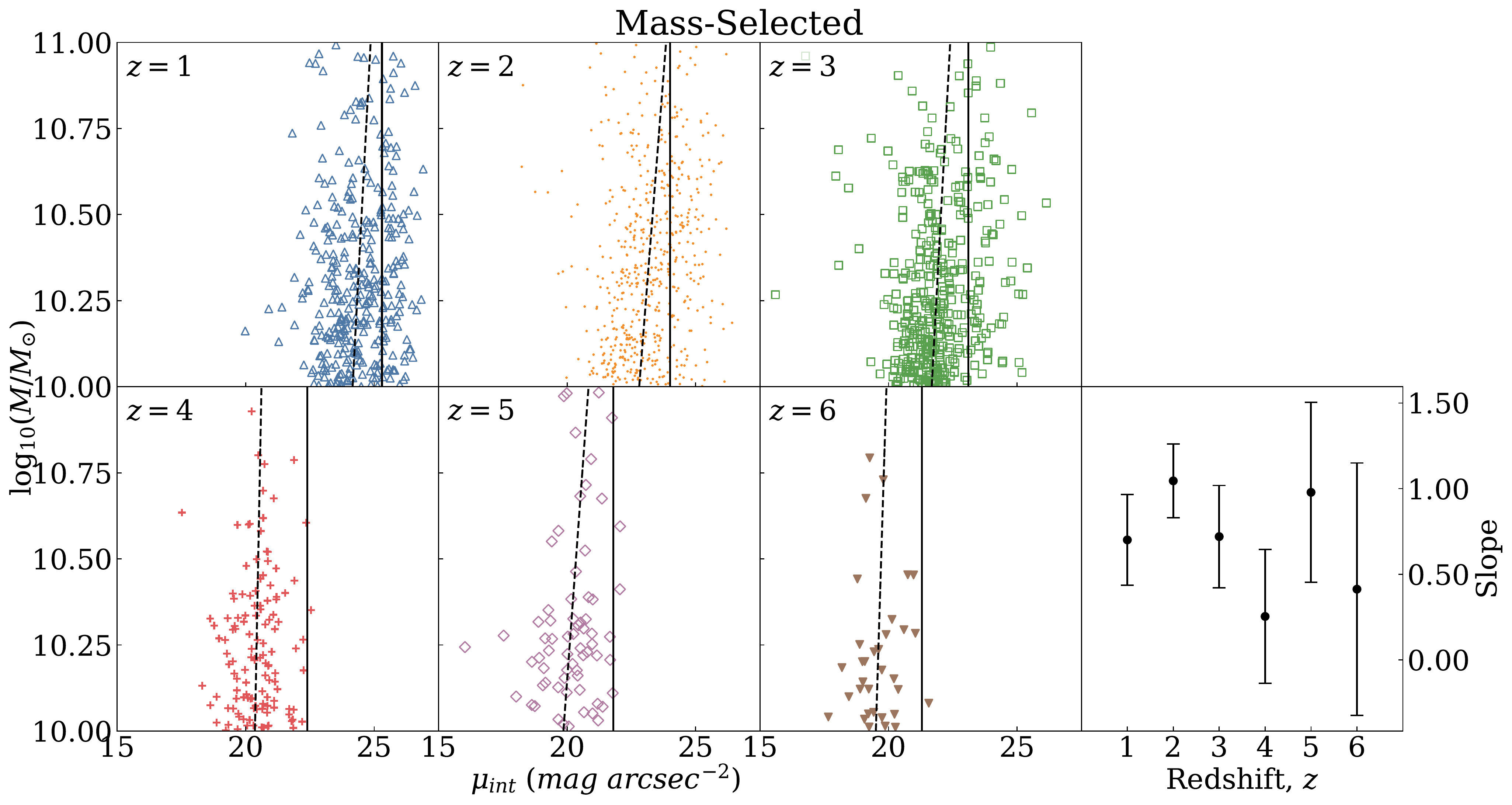}
    \caption{Relationship between the intrinsic surface brightness and galaxy stellar mass for each redshift bin. Each bin is shown in a separate panel and the final panel shows the slope of the fit for each bin. The fits are shown as dashed lines in each panel and the surface brightness completeness limit at each redshift is indicated by the solid vertical line. We find very little dependence of surface brightness on galaxy stellar mass at all redshifts, suggesting that a galaxy of any given mass could have a range of surface brightness. }
    \label{fig:masssb}
\end{figure*}

Firstly, we determine the relationship between intrinsic surface brightness and the stellar mass of mass-selected sample of galaxies for each redshift bin, as shown in Figure \ref{fig:masssb}. Each panel shows the relationship for each redshift bin with the final panel giving the evolution of the slope of the relationship. The fits for each redshift are shown as dashed lines and the surface brightness completeness limit described in \S \ref{method:sb} is shown by a solid line. We find a slight dependence of surface brightness on galaxy stellar mass whereby lower surface brightness galaxies have a higher mass however this is a very small dependence. In general, a galaxy of any given mass could have a range of surface brightness at all redshifts. This suggests that the stellar mass of a galaxy does not heavily influence the evolution of its intrinsic surface brightness. It also implies that by only reaching a certain surface brightness limit we are not solely missing low-mass galaxies, but high-mass galaxies are missing too. This has important consequences for understanding the fact that we are missing galaxies at high redshift under the surface brightness completeness limits.

\begin{figure}
\includegraphics[width = 0.475\textwidth]{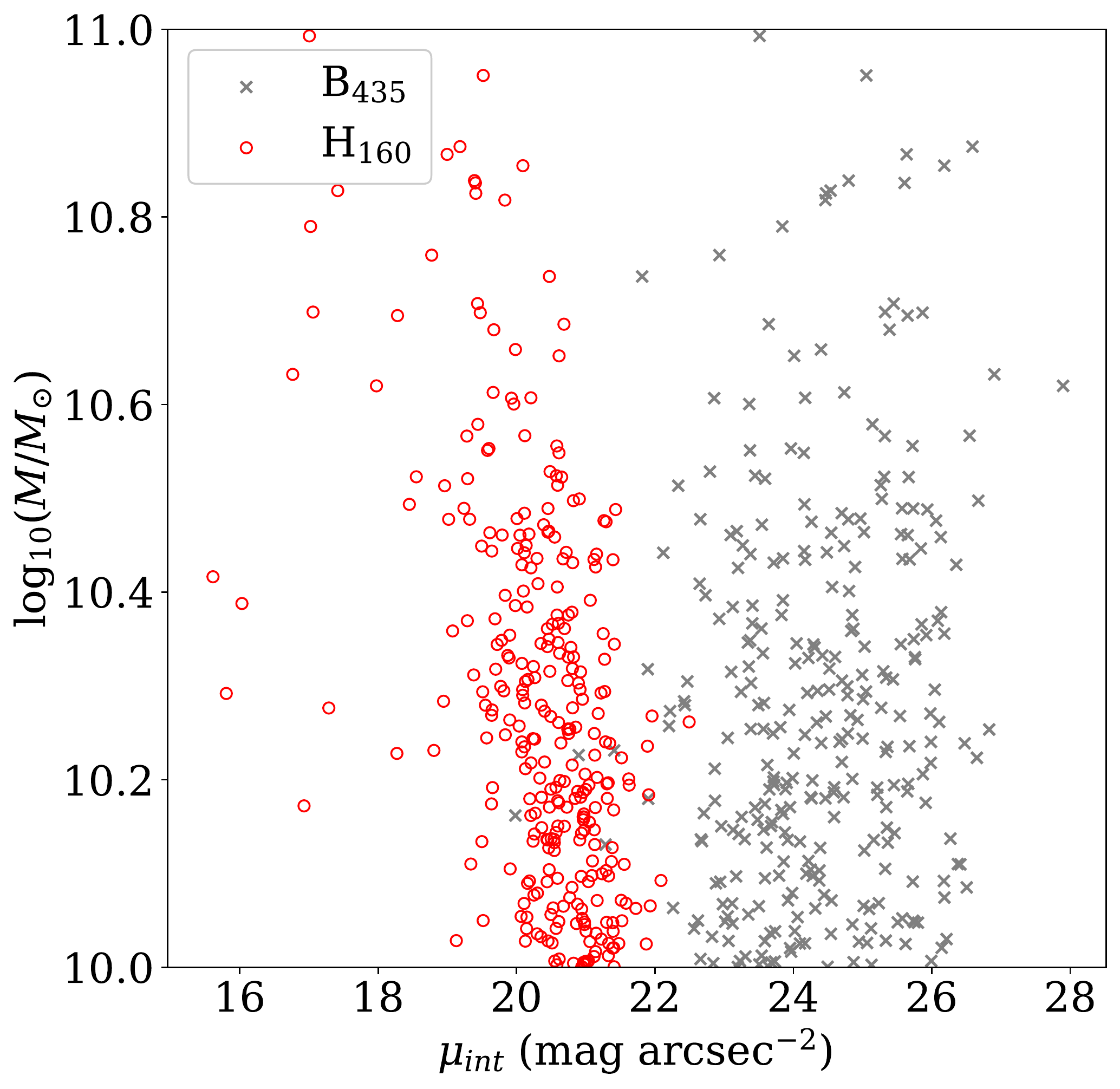}
    \caption{Intrinsic surface brightness versus stellar mass for $z = 1$ galaxies as observed with HST. The UV rest frame (B$_{435}$) is shown as grey crosses and the same measurement for the H$_{160}$ band is shown as red circles.}
    \label{fig:masssb_z1}
\end{figure}

As we probe the universe with deeper with observations, we are are able to detect galaxies with lower surface brightness. However, as we are only looking at objects in the UV rest frame at the highest redshifts with HST, we can conclude that we are missing galaxies at all stellar masses. To highlight this, we compare the relationship between the stellar mass and intrinsic surface brightness measured in the UV rest frame (B$_{435}$) and the H$_{160}$ band, as shown in Figure \ref{fig:masssb_z1}. The slope for the $H_{160}$ band is negative, as expected, as opposed to positive as seen for the $B_435$ band.  This is such that within the H-band those galaxies with the highest masses exhibit a higher surface brightness.  We do not witnesses this when observing galaxies in the rest-frame UV, there there is a large scatter and no obvious trend. From this we can conclude that it is thus likely that observations are missing low surface brightness galaxies in the UV which span all stellar masses. 

\begin{figure*}
\centering
\begin{tabular}{cc}
\subfloat{\includegraphics[width = .49\textwidth]{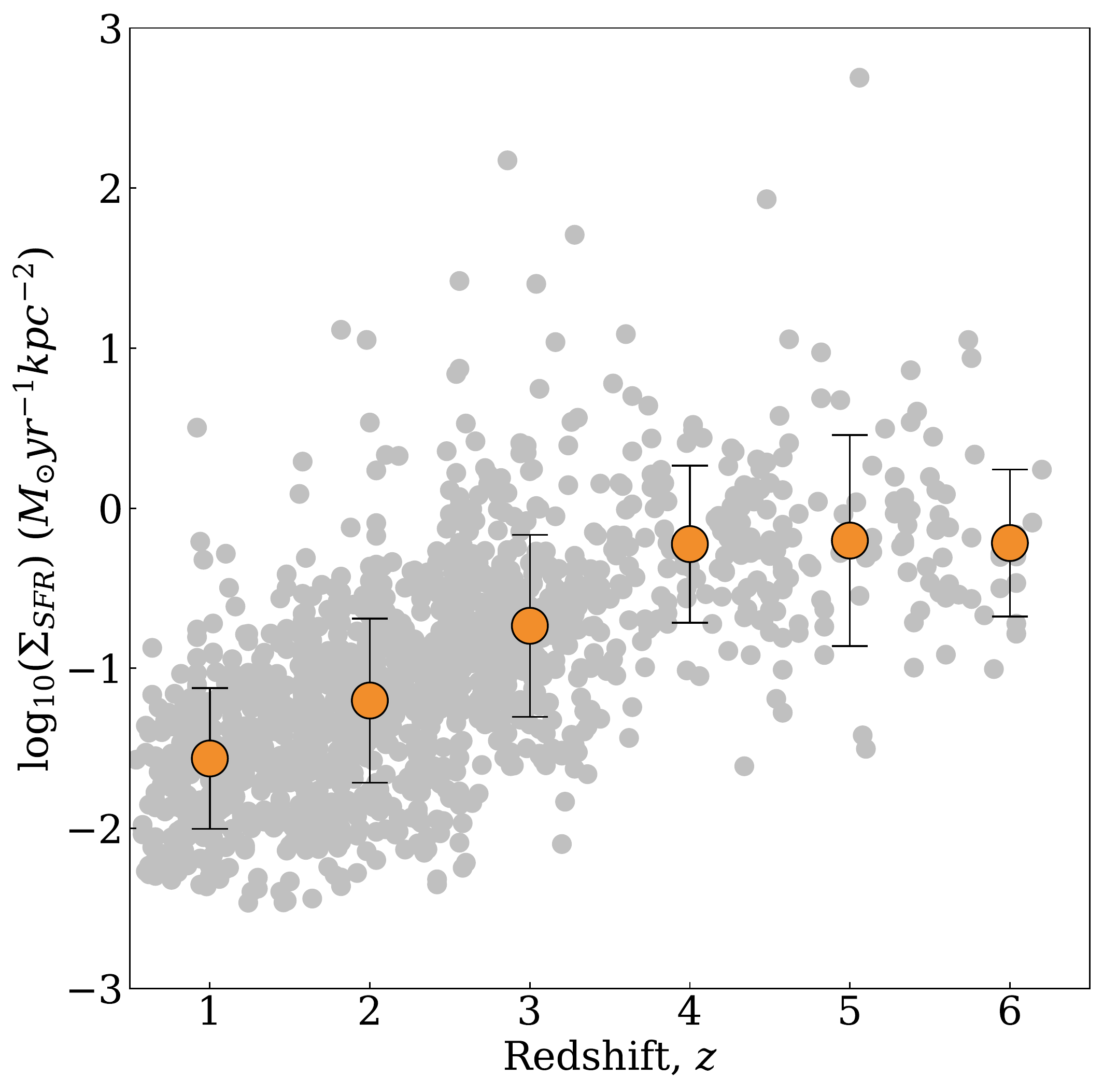}} &
\subfloat{\includegraphics[width = .49\textwidth]{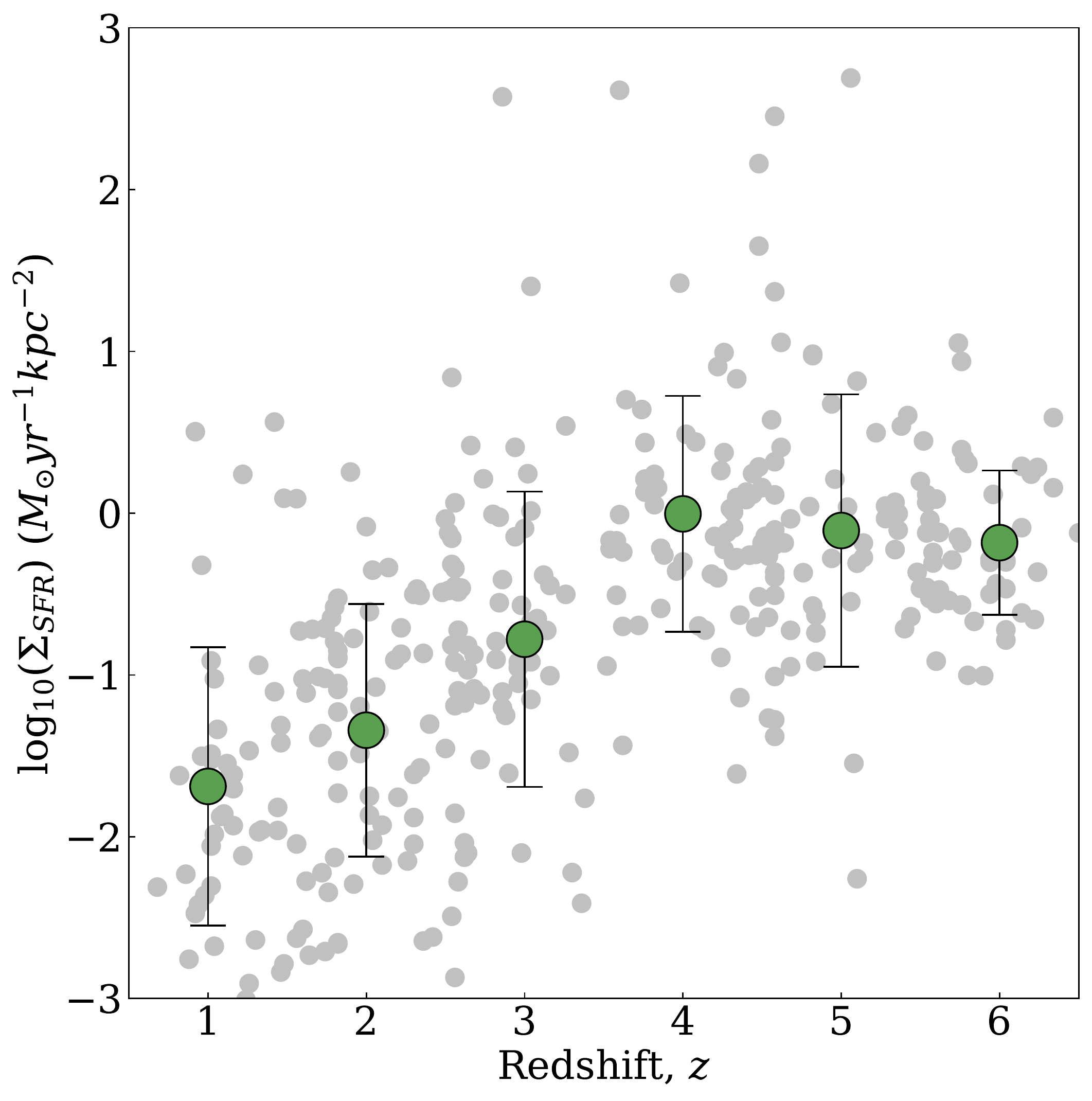}}
\end{tabular}
\caption{Left: evolution of the logarithm of the star formation rate density, $\Sigma_{SFR}$ for the mass-selected sample. Individual points are shown as grey circles. The orange circles show the mean $\Sigma_{SFR}$ for each redshift bin. The error bars on these points are given as 1$\sigma$ from the mean. Right: evolution of the logarithm of the star formation rate density for the number density-selected sample. Green circles show the mean star formation rate density for each redshift. The error bars on these points are given as 1$\sigma$ from the mean.}
\label{fig:sfrdz}
\end{figure*}

We also examine the evolution star formation rate density, $\Sigma_{SFR}$ of the two samples. $\Sigma_{SFR}$ is defined as the star formation rate per unit area where the area used is the region bound by a circle of radius $R_{\textup{Petr}}(\eta = 0.2)$. In the left panel of Figure \ref{fig:sfrdz}, we show the mean star formation rate density for the mass-selected sample as orange circles. In the right panel of Figure \ref{fig:sfrdz}, we show the same relation but for the number density-selected sample. For both samples, we see decreases in $\textup{log}_{10}(\Sigma_{SFR})$ of $1.4 \pm 0.6 M_{\odot}$yr$^{-1}$kpc$^{-2}$ and $1.5 \pm 1.0 M_{\odot}$yr$^{-1}$kpc$^{-2}$ for the mass-selected and number density-selected samples respectively.  Both results show that a higher surface brightness correlates with a larger SFR and a larger SFR per unit mass.

\begin{figure*}
    \centering
    \begin{tabular}{c}
    \subfloat{\includegraphics[width = \textwidth]{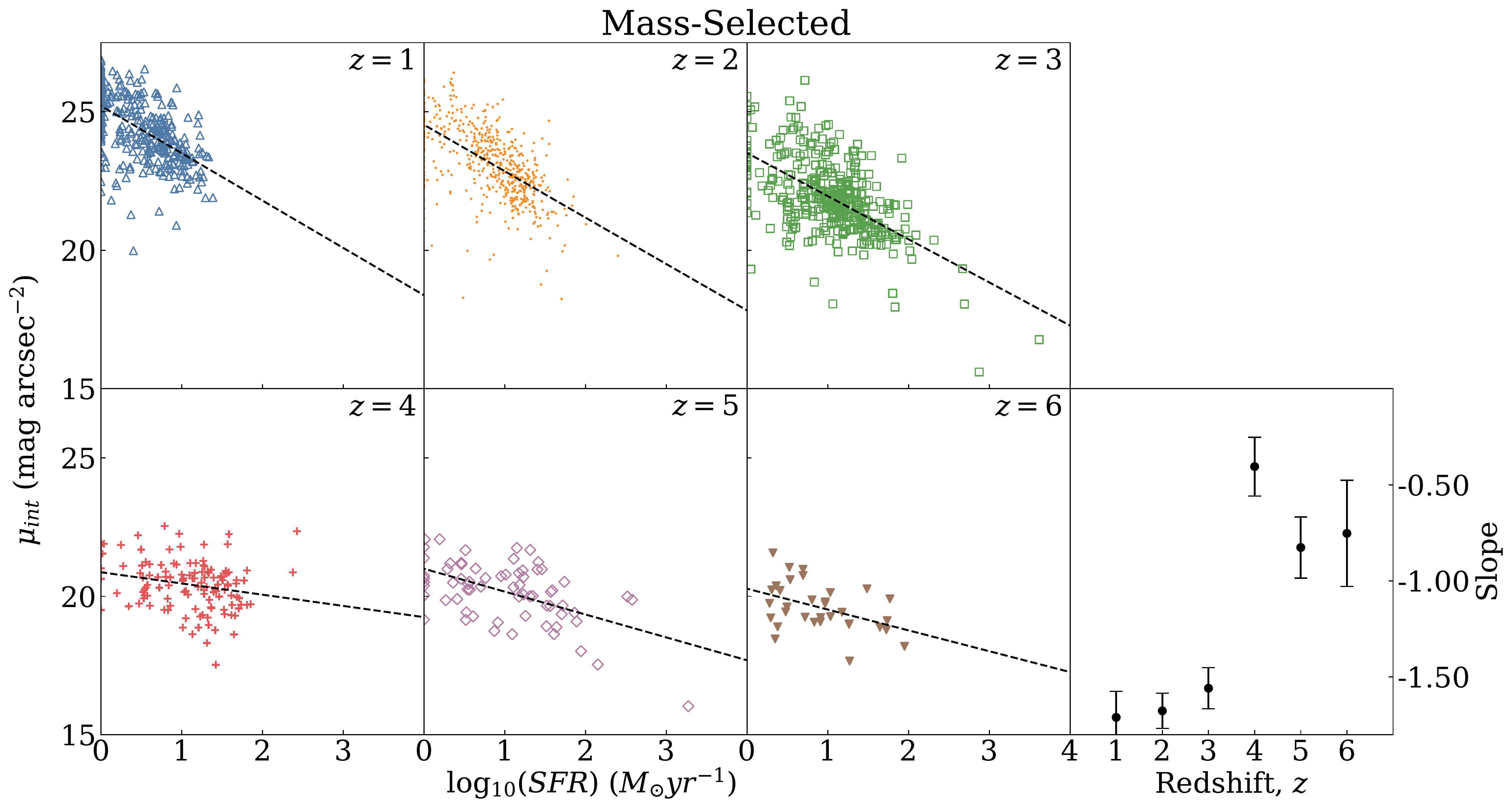}} \\
    \subfloat{\includegraphics[width = \textwidth]{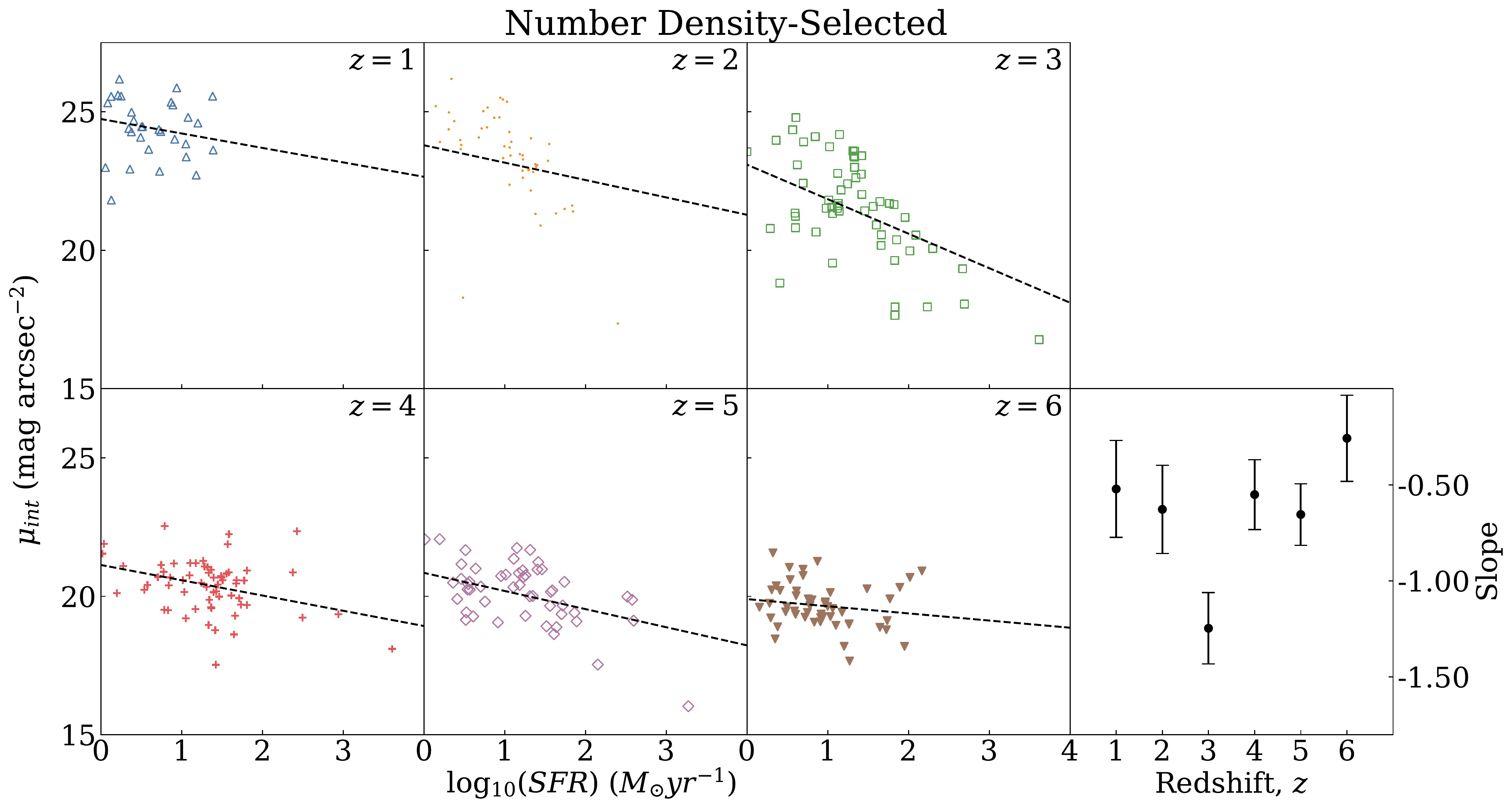}} \\
    \end{tabular}
    \caption{Top: Star formation rate versus intrinsic surface brightness for the mass-selected sample. Bottom: Star formation rate versus intrinsic surface brightness for the number density-selected sample. Each redshift bin is given as a separate panel and a straight line fit (dashed line) to the fits is given for each. The final panels give the evolution of the slope of these fits. The error bars on these points are 1$\sigma$ from the mean. For the mass-selected sample, the slope of the fits for each redshift bin gets steeper with time, on average, suggesting the dependence of surface brightness on star formation rate gets stronger with time. For the number density-selected sample, the slope remains roughly constant with time suggesting that, for this sample, there is little dependence on the relationship between star formation rate and surface brightness with time.}
    \label{fig:sfr_sb}
\end{figure*}

The relationship between the star formation rate (SFR) and the intrinsic surface brightness is shown in Figure \ref{fig:sfr_sb} for the seven redshift bins. On top we show the results for the mass selected sample and the bottom shows the same results for the number density-selected sample. Each panel shows the relationship between the two variables for each redshift bin, along with a fit to the data. The final panel shows the evolution of the slope of these fits with redshift. On average, the slope gets steeper with time for the mass-selected sample. The slope for the number density-selected sample however remains approximately constant with time. 

\begin{figure*}
    \centering
    \begin{tabular}{c}
    \subfloat{\includegraphics[width = \textwidth]{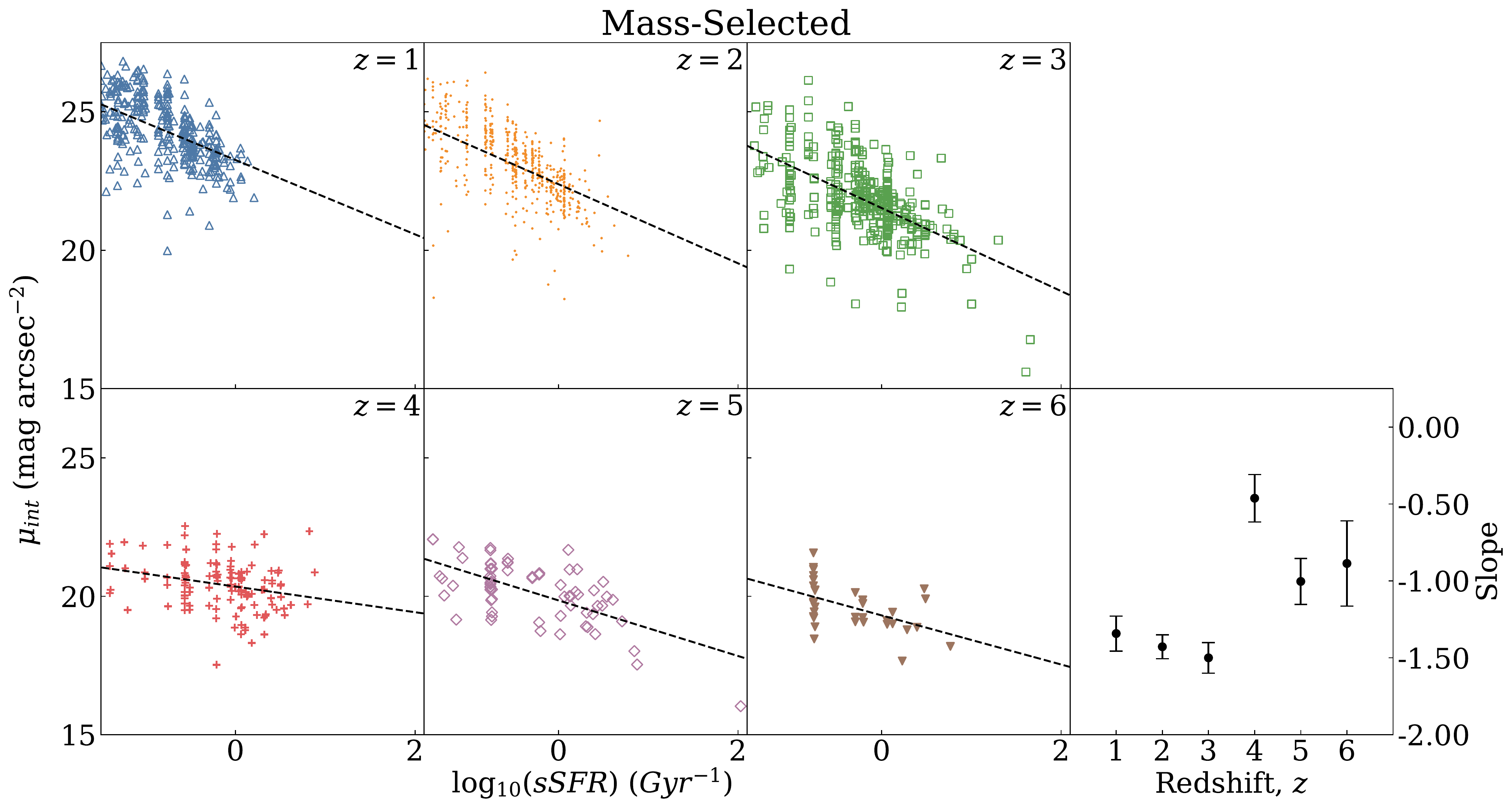}} \\
    \subfloat{\includegraphics[width = \textwidth]{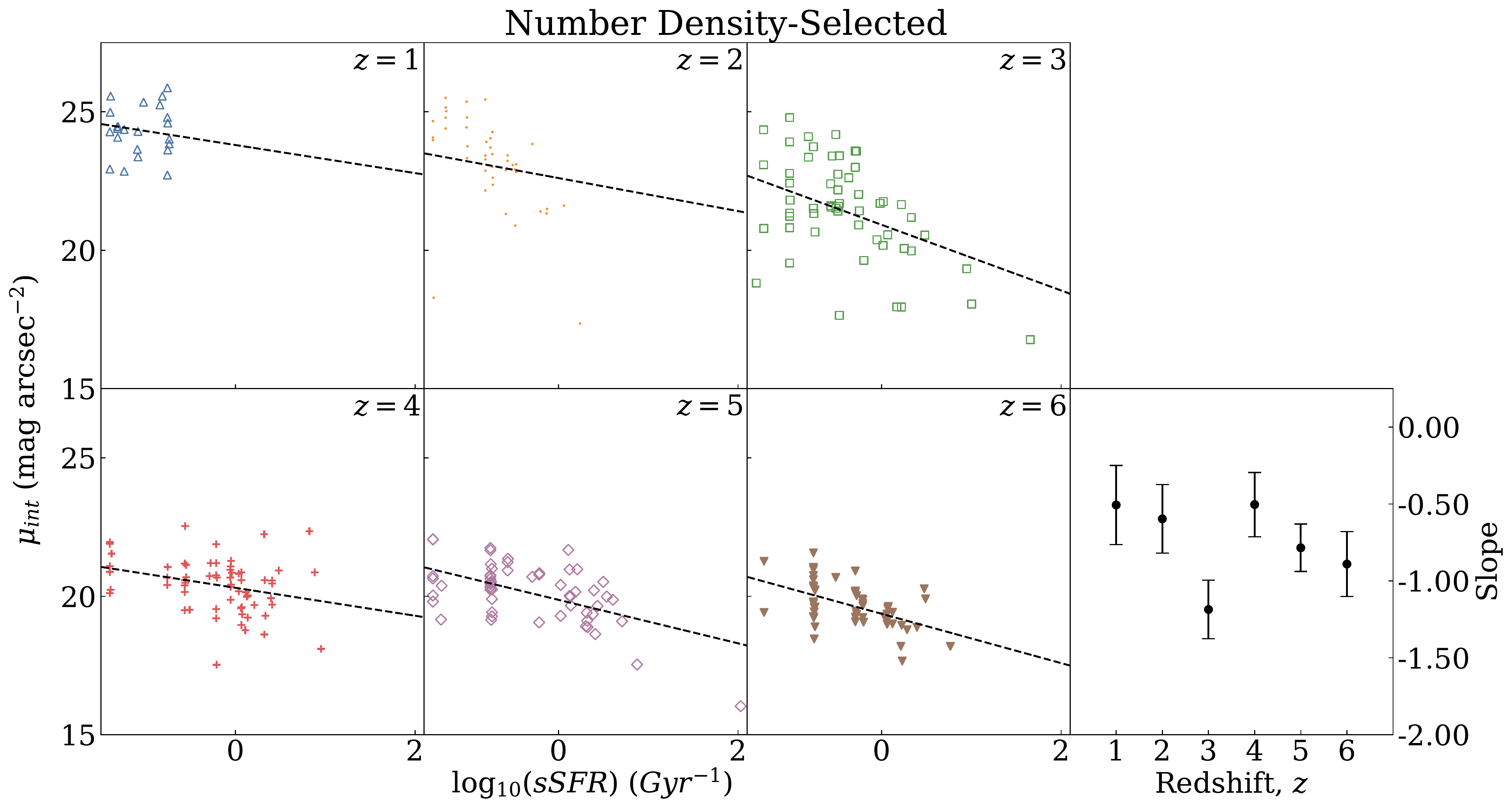}} \\
    \end{tabular}
    \caption{Top: Specific star formation rate versus intrinsic surface brightness for the mass-selected sample. Bottom: Specific star formation rate versus intrinsic surface brightness for the number density-selected sample. Each redshift bin is given as a separate panel and a straight line fit (dashed line) to the fits is given for each. The final panels give the evolution of the slope of these fits. The error bars on these points are 1$\sigma$ from the mean.}
    \label{fig:ssfr_sb}
\end{figure*}

We also show the relationship between the intrinsic surface brightness and specific star formation rate (sSFR, defined as the star formation rate per unit mass) for both samples in Figure \ref{fig:ssfr_sb}. As for Figure \ref{fig:sfr_sb}, on the top panel we show the relationship for the mass-selected sample and on the bottom we show the relationship for the number density-selected sample. Each panel shows the relationship between SFR and intrinsic SB for each of the seven redshift bins and the bottom right panel shows the evolution of the slope with redshift. The slopes of both samples get steeper with time however the mass-selected sample changes by a greater amount.  

\section{Discussion} \label{sec:discussion}

In this section we discuss our results, including what the meaning of the surface brightness evolution implies for detection of missing galaxies at high redshift, as well as for what the correlation of SB With other parameters means for the origin of the changes in observed intrinsic surface brightness.

\subsection{Surface Brightness Evolution}

The observed surface brightness evolution for both the mass and number density samples decreases with time by 0.7 $\pm$ 1.5 mag arcsec$^{-2}$ with evolution of the form $(1+z)^{-0.03 \pm 0.01}$ for the mass-selected sample and decreases by 1.2 $\pm$ 1.4 mag arcsec$^{-2}$ with evolution of the form $(1+z)^{-0.04 \pm 0.01}$ for the number density-selected sample. This suggests that the observed surface brightness remains almost constant with a slight decrease at low redshifts, which is certainly due to the more sensitive optical filters on HST.

However, that is the observed surface brightness evolution, and what we really want to study and understand is the intrinsic evolution of surface brightness once it has been corrected for redshift effects, what we call the intrinsic surface brightness, $\mu_{\textup{int}}$.  The change in the mean $\mu_{\textup{int}}$ from $z = 6$ to $z = 1$ for the mass selected-sample is 4.7 $\pm$ 1.5 mag arcsec$^{-2}$. The intrinsic surface brightness evolution of the number density-selected sample changes by 5.0 $\pm$ 1.4 mag arcsec$^{-2}$ over the same range. This is a very large change in surface brightness over this redshift range, where $\mu_{\textup{int}}$ decreases by 0.5-0.7 mag arcsec$^{-2}$ per redshift interval. 

This is consistent with previous studies using both ground based and \textit{HST} observations that find an evolution of 1-2 mag between $z = 0$ and $z = 1$ \citep[e.g.][]{schade96, lilly98, labbe03, barden05}, but now extended to much higher redshifts. This change in intrinsic surface brightness is likely due to the fading of sources or an intrinsic luminosity difference.   The question is - what is producing this change in intrinsic surface brightness evolution, and what does it imply for our detectability for distant galaxies? To answer this we consider the effects of size, dust, star formation and detection limits in understand these issues.

\subsubsection{Effects of UV Luminosity Evolution}

As we are examining the evolution of the UV luminosity surface brightness, one area where we must first try to understand the evolution of the SB is within the UV luminosity itself. As explained in \S \ref{sec:arg}, we are able to trace the evolution of this UV luminosity function, and more importantly, the value of M$^{*}_{\rm UV}$ and how it evolves with time.  Although the total SFR decreases at $z > 2$ overall per unit volume, the UV flux of individual galaxies increases, as characterized by the UV luminosity function.  Using this data we know that the $z = 2$ characteristic magnitude, M$^*_{UV} = -20.33\pm0.50$ \citep{arnouts05} and at higher redshifts this becomes brighter to $-20.94\pm0.20$ \citep{bouwens15}, as discussed in \S \ref{sec:method_arg}.Thus a representative amount of evolution is for galaxies to become fainter, on average, by about 0.61 magnitudes over this epoch.  This is certainly much less than the 4 to 5 magnitudes of observed SB evolution.  Therefore the intrinsic SB evolution cannot be accounted for solely or even primarily by an evolution in UV flux.

\subsubsection{Size Corrections}

The size evolution of galaxies is well documented \citep[e.g.][]{trujillo07, buitrago08, dokkum08, cassata10, whitney19}. Because galaxies evolve in size to become larger at lower redshifts, this will in principle act to decrease the surface brightness of galaxies.  Thus we test to determine whether this observed size evolution is causing an 'artificial' evolution in the surface brightness. To determine this we correct both the observed and intrinsic surface brightness values by setting the size of all galaxies to that of the median size at $z = 6$ as measured in \cite{whitney19}. 
In all cases (for both samples and for both measures), we find that the fit to the evolution becomes less steep, and the change in intrinsic SB decreases from $\Delta\mu_{1\rightarrow6}$ = 4.7 $\pm$ 1.5 mag arcsec$^{-2}$ to $\Delta\mu_{1\rightarrow6}$ = 3.5 $\pm$ 1.8 mag arcsec$^{-2}$ for the mass-selected sample and decreases from $\Delta\mu_{1\rightarrow6}$ = 5.0 $\pm$ 1.4 mag arcsec$^{-2}$ to $\Delta\mu_{1\rightarrow6}$ = 3.9 $\pm$ 1.9 mag arcsec$^{-2}$ for the number density-selected sample. Therefore, the value of mag arcsec$^{-2}$ changes by $\sim$1 mag arcsec$^{-2}$ for both samples. This suggests that whilst the size evolution is causing some of the evolution we see in the surface brightness, it is not the primary cause.  When we account for the size and luminosity evolution, this still leaves $\sim$ 3.5 magnitudes of SB unaccounted for that must be due to other effects.

By correcting the surface brightness using eq. \ref{eq:sbint_z6}, we are effectively examining the absolute magnitude at a fixed size. Ultimately we find that there is a small difference between the size corrected and uncorrected surface brightness evolution, indicating that the effect of size evolution and surface brightness evolution on incompleteness is small. If the surface brightness were the main contributor to the incompleteness, low surface brightness galaxies would be outnumbered by high-surface brightness galaxies. Our results seem to indicate this is not what is happening. Instead, our findings seem to be a result of the intrinsic faintness of the rest-frame UV magnitudes of all galaxies, including massive systems.
 
\subsection{Effects of Dust} \label{sec:dust}

Here we explore the possibility of dust extinction producing an apparent decrease in the surface brightness from $z = 6$ to $z = 1$. In this work we use the UV rest-frame and as such, the light is significantly affected by dust attenuation. 

We calculate the estimated dust extinction using the relation between dust extinction and the UV-continuum slope $\beta$ from \cite{meurer99}:

\begin{equation}
    A_{1600} = 4.43 + 1.99\beta
\end{equation}

\noindent where $A_{1600}$ is the dust extinction at the rest-frame wavelength $\lambda = 1600$\AA. In the case where this relation implies a negative extinction, $A_{1600}$ is assumed to be 0. Typically, higher redshift objects are found to have bluer UV-continuum slopes than lower redshift objects, implying the presence of younger stellar populations and lower metallicities in higher redshift objects \citep{wilkins13}. A smaller value of $\beta$ therefore suggests there is less dust extinction at high-$z$ \citep{meurer99, lehnert03, bouwens09}.  This implies that as dust grows in importance with time it will have more of an effect on the measured UV luminosity, which will therefore also produce a change in the SB of the galaxies. 

To determine the effect of this we consider the measured extinction for $z \geq 3$ galaxies due to the fact the $\beta$ is a measure of evolved populations and as such, the high values of $\beta$ measured at low redshift are artificially inflated by the abundance of early-type galaxies. We find that for the mass-selected sample, there is an increase of 0.1 $\pm$ 3.4 mag between $z = 6$ and $z = 3$. When fitting the values of $A_{1600}$ to a straight line, we calculate a best-fit slope of -0.15 $\pm$ 3.4, so on average, there is an increase in the extinction but it is not significant. The value of $A_{1600}$ for the number density-selected sample increases by 2.6 $\pm$ 4.3 mag from $z = 6$ to $z = 3$. The dust extinction therefore plays a small role in the decrease in intrinsic surface brightness, with $\sim$0.1 mag being accounted for in the mass-selected sample and $\sim$2.6 mag being accounted for in the number density-selected sample when considering the redshift range $3 \leq z \leq 6$. After considering size and dust corrections, we are left with $\sim$3.4 mag arcsec$^{-2}$ of SB evolution unaccounted for within the mass-selected sample and  $\sim$1.3 mag arcsec$^{-2}$ unaccounted for within the number-density selected sample.

\subsection{UV Magnitude Evolution Modelled}

\begin{figure}
\centering
\includegraphics[width = 0.475\textwidth]{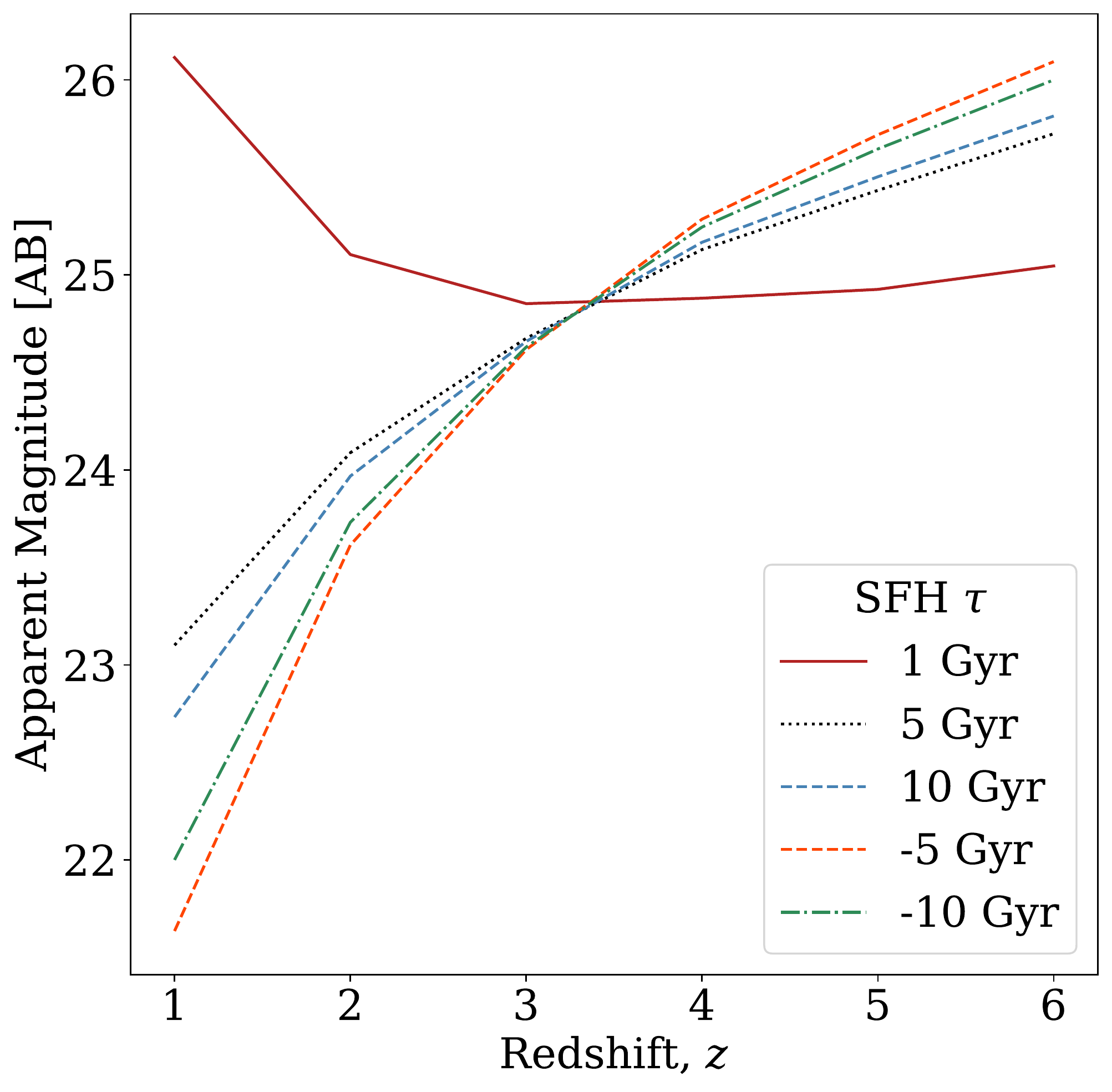}
    \caption{Evolution of the apparent magnitude for five different star formation histories. The SED is convolved at each redshift with the filter that corresponds to the UV rest-frame wavelength at that redshift. We use three decreasing star formation histories; $\tau$ = 1 Gyr is shown as a solid red line, $\tau$ = 5 Gyr is shown as a dotted black line and $\tau$ = 10 Gyr is shown as a dashed blue line. We also show two examples of increasing star formation histories; $\tau$ = -5 is shown as a dashed orange line and $\tau$ = -10 is shown as a green dashed line. There is a significant difference between $\left|\tau\right|$ = 1 Gyr and $\left|\tau\right|$ $>$ 1 Gyr whereby the UV rest-frame magnitude decreases in brightness from $z = 6$ to $z = 1$ for $\tau$ = 1 Gyr but increases in brightness for the other scenarios. The rising star formation histories (negative $\tau$) yield a greater increase in magnitude than for the decreasing star formation histories.  These star formation histories all give an evolution in SB that is opposite to what we observe. The magnitude values on the y-axis are not representative of the true values however the change in magnitude, which depends on mass, but the relative change is accurate for all systems.}
    \label{fig:modelmag}
\end{figure}

We model the apparent magnitude evolution for the redshift range $1 \leq z \leq 6$ for a number of different star formation histories using \textbf{SMpy}\footnote{https://github.com/dunkenj/smpy} \citep{duncan14}. The SED is convolved at each redshift with the filter corresponding to the UV rest-frame wavelength at that redshift. The filters are given in Table \ref{tab:imgprocessing}. We use the same simple stellar population models and initial mass function used when completing the stellar mass fitting as described in \S \ref{sec:mass}. We model the evolution of the rest-frame magnitude using a star formation history given by 

\begin{equation}
    SFR \propto e^{\frac{-t}{\tau}}
\end{equation}

\noindent and vary the value of $\tau$. The evolution of the magnitude using star formation histories with $\tau$ = -10 Gyr, -5 Gyr, 1 Gyr, 5 Gyr, and 10 Gyr are  as shown in Figure \ref{fig:modelmag} where $\tau$ = -10 Gyr is shown as a green dashed line, $\tau$ = -5 Gyr is shown as an orange dashed line, $\tau$ = 1 Gyr is shown as a red solid line, $\tau$ = 5 Gyr is shown as a black dotted line and $\tau$ = 10 Gyr is shown as a blue dashed line. The magnitude values on the y-axis are not representative of the true value as this is mass dependent, but the change in magnitude modelled is accurate for all systems. There is a significant difference between the scenarios where the star formation histories are decreasing with $\tau$ = 1 Gyr and $\tau$ $>$ 1 Gyr whereby galaxies appear to get dimmer with the first scenario and brighter with the second. The first scenario is where the star formation occurs very quickly during the early universe, whereby the others are for a more gradual star formation that continue. The scenarios where we model a rising star formation history yield a change in magnitude similar to that of $\tau$ $>$ 1.  However the change is much larger in these cases and therefore do not match well with the trend in surface brightness we see. For star forming galaxies we would expect these $\tau$ values to be somewhat larger, and only near $\tau \sim 1$ at $z < 3$ for the most massive systems that undergo most of their star formation early in the universe's history. The model we use assumes a dust extinction of $A_V = 0$ and a metallicity of $Z = Z_{\odot}$. We also assume these values are constant with redshift therefore this figure only illustrates the age effects on the magnitude evolution but in reality, the dust extinction increases with time, as explained in \S \ref{sec:dust}, and also scales with metallicity so the evolution may be steeper than seen here. 

If we assume a star formation history with $\tau$ = 1 Gyr, we see a decrease of 1.1 mag from $z = 6$ to $z = 1$ which would partially account for the evolution seen in the surface brightness not already accounted for by size and dust. However, our systems are known to be star forming over a long period of time, and therefore the $\tau$ value would be larger. If we take the information from empirical measurements of the value of $\tau$ \citep[e.g.][]{ownsworth16} we find that $\tau \sim 2-5$ Gyr.  If we assume that our galaxies have a similar $\tau$ which is the case for systems at these mass ranges at $z < 3$, then we would find that the star formation history would {\em increase} the SB by a few magnitudes. This would only exacerbate the problem of understanding the decline of the SB for these galaxies at lower redshifts. We have nonetheless already accounted for the decline in the UV brightness from using the luminosity function changes from $z \sim 6$ to lower redshifts. Thus we can conclude from this section that the decline in surface brightness we observe in our selections is not due to an evolution in the star formation rate.  If anything these estimates are the maximum change in the brightness as the star formation history is known to increase from $z = 6$ to $z = 3$ which would have the effect of only increasing further the brightness of our sample.

\subsubsection{Image Simulations - How many galaxies are we missing?}

By artificially redshifting a sample of galaxies, we are able to determine whether the same galaxies we see at low redshift are detectable at high redshift. As a fiducial experiment we simulate the galaxies in our $z = 2$ sample to $z = 6$ and determine how many of these we would detect, at what S/N, and at what measured surface brightness, as described in \S \ref{sec:arg}.

When we do this experiment we find that only $\sim$16\% of these simulated $z = 2$ galaxies are still detectable when simulated to be at $z = 6$ (at a $S/N > 5$), compared to 94\% of the original galaxies at $z = 2$ with a $S/N > 5$, with the remaining 6\% of detected galaxies at a $S/N < 5$ down to our magnitude limit. This suggests we are not detecting a significant number of galaxies at these high redshifts due to their observed low surface brightness produced by the effects of redshift. 

The characteristic luminosity ($L^*$), as determined from the characteristic magnitudes given by \cite{arnouts05} and \cite{bouwens15}, decreases by a factor of 1.75 from $z = 6$ to $z = 2$. By calculating the equivalent change in luminosity of the redshifted sources using 

\begin{equation}
    \frac{L_{z=6}}{L_{z=2}} = 10^{0.4\cdot\Delta M}, 
\end{equation}

\noindent we measure that the intrinsic luminosity of all galaxies artificially redshifted to $z = 6$ is 5.75 times brighter in SB than that of the original galaxies at $z = 2$. Therefore, the evolution we see in the artificially redshifted galaxies is greater than that of the characteristic luminosity. This strongly implies that the galaxies at $z = 2$ are not the descendants of the galaxies at $z = 6$ just by examining how much evolution they would have undergone.

When artificially redshifting the sample of $z = 2$ galaxies, we find that the change in surface brightness, both observed and intrinsic, does not correlate with the change we see in the real galaxies. For the observed surface brightness, we see a decrease in surface brightness of 1.0 $\pm$ 1.9 mag arcsec$^{-2}$ when comparing the redshifted $z = 6$ galaxies (evolved with the luminosity function correction) to the original $z = 2$ galaxies, as shown in Figure \ref{fig:simsb}. For the real sample, we see an increase in surface brightness of 0.7 $\pm$ 1.1 mag arcsec$^{-2}$ over the same redshift range. For the intrinsic surface brightness, we see a change of 1.9 $\pm$ 1.5 mag arcsec$^{-2}$ in the simulated sample whereas for the real sample, there is a change of 3.5 $\pm$ 1.6 mag arcsec$^{-2}$ over the same redshift range. 

We conclude from this analysis that there are a significant number of missing galaxies at high redshifts that are driving the intrinsic average surface brightness evolution. Because it is only the brightest galaxies we are detecting at $z \sim 6$ due to our SB limits, this drives up the intrinsic measurement of the average surface brightness. This is true even when examining a mass- or number density-selected sample. It is thus not simply due to missing lower mass galaxies. In fact, as our simulations show we are missing 84\% of the $z = 2$ galaxies when they are simulated to $z = 6$ even when we take into account the average increase in magnitude from the UV luminosity function.  From this we conclude that there is a significant population of galaxies that remain undetected at these higher redshifts \citep[e.g.][]{conselice16}. Most of these galaxies are likely to be span our entire mass range, and not just be low-mass systems. Therefore, our previously stated mass completeness limit underestimates selection effects.

The density of our detected galaxies (those with S/N $>$ 5) at $z = 2$ is measured as $3.2\times10^{-3}$Mpc$^{-3}$. The density of our detected galaxies at $z = 6$ is almost a factor of 10 smaller at $2.5\times10^{-4}$Mpc$^{-3}$. The measured density of the simulated $z = 6$ galaxies (evolvoed with the luminosity function) is $4.1\times10^{-3}$Mpc$^{-3}$. This is 1.6 times greater than the real $z = 6$ galaxies. This suggests we are missing a significant number of high redshift galaxies in the real observations. The fact that there are undetected galaxies at high redshift is consistent with the actual number of galaxies being higher than the number we can actual see \citep{conselice16}.  Many of these missing galaxies will be lower mass, but a significant fraction will be high mass systems.  

\subsection{The origin of high SB galaxies}

In this section we discuss how to measure how the SB evolves intrinsically and what this implies for the evolution of galaxies in terms of their star formation rate and gas densities that produce this star formation. One issue is that it is clear that at the highest redshifts we are only observing galaxies at the highest redshifts with the very highest intrinsic SB levels. These are much higher in SB than galaxies in the nearby universe and the question is what is the origin of these systems and how do they relate to lower redshift galaxies?

\subsubsection{Star Formation Rate Density}

For both mass and number density selected samples, we find that on average, there is a decrease in the star formation rate density ($\Sigma_{SFR}$) over time. The gas density and fraction are observed to be greater at higher redshifts \citep{tacconi13, gonzalez17} which leads to a higher $\Sigma_{SFR}$ by extrapolating from the Kennicutt-Schmidt law \citep{barden05, mosleh12}. Increased size at low redshift also contributes to the lower $\Sigma_{SFR}$ we see at low redshift. However, from the change we see in the star formation rate over redshift (Figure \ref{fig:sfrdz}), this size evolution is unlikely to be a significant cause of the decrease in $\Sigma_{SFR}$ at lower redshifts.

We find a difference in $\Sigma_{SFR}$ of 1.4 $\pm$ 0.6 M$_{\odot}$yr$^{-1}$kpc$^{-2}$ for the mass-selected sample and a difference of 1.5 $\pm$ 1.0 M$_{\odot}$yr$^{-1}$kpc$^{-2}$ for the number density-selected sample over the redshift range $1 < z < 6$. This suggests that the evolution in luminosity cannot be solely driven by the star formation rate density. This reinforces our conclusions from the previous argument that there are multiple factors contributing to the origin of the 4 to 5 mags of evolution in intrinsic SB observed.

\subsubsection{Star Formation Rate and Specific Star Formation Rate}

As we have shown, the surface brightness of a galaxy depends strongly on the star formation rate, particularly in the UV. In the case of the mass-selected sample, this relationship appears to grow stronger as time progresses with galaxies that have a high star formation rate generally exhibiting a brighter SB. In the case of the number density selected sample the slope of the relation between these quantities remains approximately constant with time suggesting that there is little dependence on the relationship between star formation rate and intrinsic surface brightness with time. 

The two samples differ in that one contains galaxies that are at the same mass and one that contains galaxies of differing mass, depending on the number density requirement with the lower redshift bins containing galaxies of a higher mass than the higher redshift bins. From this, we can infer that the surface brightness of galaxies at high redshift of a given mass depends less on the star formation rate than galaxies of the same mass at lower redshifts. On the other hand, if we directly track galaxies through time, we infer that the relationship between star formation rate and surface brightness does not change as the galaxies evolve and grow in mass. 

The evolution of the relationship between the specific star formation rate and intrinsic surface brightness is similar to that between the SFR and $\mu_{int}$ whereby the mass-selected sample shows a stronger dependence between the two at lower redshifts and for the number density-selected sample remains roughly constant with time. 

\section{Conclusions} \label{sec:summ}

We present an analysis of the surface brightness evolution of two separate samples (a mass-selected sample and a number density-selected sample) of galaxies taken from the GOODS-North and GOODS-South fields of the Cosmic Assembly Near-infrared Deep Extragalactic Legacy Survey. We examine UV rest-frame ($\lambda \sim 2000\AA$) images and find that strong evidence for surface brightness evolution whereby on average, galaxies get intrinsically brighter per unit area, with time. This is the case for both the mass-selected and number-density samples. The evolution is consistent with the form given by $\propto (1+z)^{-0.18 \pm 0.01}$ and $\propto (1+z)^{-0.19 \pm 0.01}$ for each sample respectively. We explore possible causes of this evolution. 

Size evolution is well known, and is now well quantified, so we test to determine whether this is producing the surface brightness evolution by setting the size of all galaxies to that of the average size of the galaxies at $z = 6$ (Whitney et al. 2019). We find that SB the evolution is not significantly changed when accounting for this, thus the increase in the size of galaxies is not producing an 'artificial' surface brightness evolution. We also find that dust extinction as a possible cause of the evolution we see in $\mu_{int}$ but we find that it plays a small role, contributing 0.1 mag of evolution between $z = 6$ and $z = 3$ for the mass-selected sample and 2.6 mag for the number density-selected sample over the same redshift range. 

We find that the star formation rate density ($\Sigma_{SFR}$) decreases with time, however the change does not completely explain the significant evolution we see in the intrinsic surface brightness.  Thus a large portion of the intrinsic evolution for SB within galaxies is left unexplained. Thus, we are see an unnatural evolution in the amount of SB changes which thus must be due to an observational bias -- we are missing high redshift galaxies in our observations, some of which are likely quite massive systems.  

The stellar mass is known to be the cause of differences in galaxy evolution and is a measure of the formation and merging history of a galaxy \citep{bhatawdekar19}. We find that the intrinsic surface brightness of galaxies in the mass-selected sample does not rely heavily on the stellar mass; for each redshift bin, the slope of the relationship between the two quantities is very close to zero. This suggests that the stellar mass of a galaxy does not correlate with a galaxy's surface brightness. It also implies that the selection in the UV at these redshifts does not give mass completeness at any mass, even for the most massive, star forming galaxies.

To further demonstrate this we artificially redshift a sample of $z = 2$ galaxies to $z = 6$ to determine the level of surface brightness and what fraction of these we would still be detected at the higher redshift. We find that the surface brightness of these redshifted galaxies is much lower than that of the real $z = 6$ galaxies suggesting that we are not detecting the low SB galaxies we see at low redshift. This remains true when we consider the amount of evolution in the UV luminosity function when carrying out these simulations. 

In the case of the mass-selected sample, the star formation rate and specific star formation rate of galaxies depend strongly on the surface brightness at low redshift however this relationship gets weaker as redshift increases. We find that the relationship between these parameters is approximately constant for the number density-selected sample so whilst there is a dependence on SFR and $\mu_{int}$, this dependence does not change with time. 

Overall, we conclude that the high surface brightness galaxies we find at high redshift are not perfect analogous to starburst galaxies seen at lower redshifts. It is also likely that there are many missing galaxies at these redshifts which will be discovered with telescopes that can probe deeper than HST, such as the forthcoming James Webb Space Telescope (JWST).   Uncovering these galaxies will require that we obtain fundamentally much deeper imaging which will be carried out with first generation imaging with JWST.  This may reveal a new population of distant high redshift galaxies that are not just lower mass, but including lower SB massive systems.  These results may in fact lead up to alter our understanding of galaxy formation and the history of star formation and mass assembly in our universe's history.  

\section{Acknowledgements}

We thank the anonymous referee for all their very useful comments and suggestions which have lead to a greatly improved paper. This work is based on observations taken by the CANDELS Multi-Cycle Treasury Program with the NASA/ESA \textit{HST}, which is operated by the Association of Universities for Research in Astronomy, Inc., under NASA contract NAS5-26555. We thank the CANDELS team for their heroic work making their products and data available. We acknowledge funding from the Science and Technology Facilities Council (STFC).




\bibliographystyle{yahapj}
\bibliography{bibliography}

\appendix

\section{Surface Brightness Dimming Derivation} \label{app:sb_evol}

As explained in \S \ref{sec:sbdim}, there are five factors of $(1+z)$ to take into consideration when determining the extent of surface brightness dimming. One term of $(1+z)$ originates from the energy reduction of a photon. When a photon is emitted, it has a wavelength $\lambda_1$ and has energy $E_{emit}$ given by 

\begin{equation}
    E_{emit} = \frac{hc}{\lambda_{emit}}.
\end{equation}

\noindent We know that observed wavelength, $\lambda_{obs}$, is a factor of $(1+z)$ larger than the emitted wavelength due to redshift, therefore, the observed energy, $E_{obs}$, is given by

\begin{equation}
    E_{obs} = \frac{hc}{\lambda_{emit}(1+z)}.
\end{equation}

\noindent From this, we can say that the energy is a factor of $(1+z)$ smaller at the time of observation than at the time of emission, giving the first factor of $(1+z)$.  Another factor is due to time dilation; two photons emitted in the same direction $\delta t_{emit}$ are separated by a proper distance $c\delta \times t_{emit}$. At observation, the proper distance is $c \times \delta t_{emit} \times (1+z)$. The two photons are then detected at a rate of 

\begin{equation}
    \delta t_{obs} = \delta t_{emit}(1+z).
\end{equation}

\noindent This yields a second factor of $(1+z)$. The two factors that are due to energy reduction and time dilation are the two factors associated with the luminosity distance $d_L$. The luminosity distance $d_{L}$ is used when determining the flux emitted from an object with luminosity $L$:

\begin{equation}
    f = \frac{L}{4\pi d_L^2}.
\end{equation}

\noindent the luminosity distance is related the the proper distance $r$ by the relation 

\begin{equation}
    d_L = r(1+z).
\end{equation}

\noindent We can therefore express the flux in terms of the luminosity distance as:

\begin{equation}
    f = \frac{L}{4\pi r^2(1+z)^2}.
    \label{eq:flux1}
\end{equation}

The two factors in the denominator are due to the energy and time dilation effects as described above.  The third and fourth factors arise from the change in angular size of the object; for an object of diameter $d$ and angular size $\delta \theta$, the angular diameter distance is given as:

\begin{equation}
    d_A = \frac{d}{\delta \theta}.
\end{equation}

\noindent From the Robertson-Walker metric the distance between the two ends of the object can be defined as 

\begin{equation}
    ds = a(t_e)S_{\kappa}(r)\delta \theta
\end{equation}

\noindent where $a(t_e)$ is the scale factor (equivalent to $(1+z)^{-1}$) and $S_{\kappa}(r)$ is some function of the proper distance that is dependent on the geometry of the Universe. We set the distance $ds$ to be the diameter of the observed object. Therefore,

\begin{equation}
    d_{A} = \frac{S_{\kappa}(r)\delta \theta}{(1+z)}.
\end{equation}

\noindent For a flat Universe, $S_{\kappa}(r)$ is equal to the proper distance $r$ so the angular diameter distance is given by

\noindent For a flat Universe, the angular diameter distance is given by

\begin{equation}
    d_A = \frac{r}{(1+z)}.
\end{equation}

\noindent Substituting this into equation \ref{eq:flux1}, we see that the flux is reduced by a factor of $(1+z)^4$ due to redshift:

\begin{equation}
    f = \frac{L}{4\pi d_A^2(1+z)^4}.
\end{equation}

The fifth and final factor arises due to the change in unit bandpass wavelength from emission to detection. A filter has central wavelength $\lambda$ and width $\Delta \lambda$. Both the observed central wavelength and the observed band width increase by a factor of $(1+z)$. For a fixed filter width this results in a reduction of observed flux from a galaxy.  The observed bandpass wavelength can therefore be given as 

\begin{equation}
    \lambda_{obs} = \lambda_{emit}(1+z) + \Delta \lambda_{emit}(1+z).
\end{equation}

\noindent From these arguments, we can see that 

\begin{equation}
    \mu_{int} \propto (1+z)^{-5}
\end{equation}

\noindent in units of $erg s^{-1} cm^{-2}$ \AA$^{-1}$, or ST magnitudes, where $\mu_{int}$ is the intrinsic surface brightness.

As shown in \S \ref{sec:sbdim}, 

\begin{equation}
    \mu_{int} \propto (1+z)^{-3}
\end{equation}

\noindent in units of $erg s^{-1} cm^{-2}$ Hz$^{-1}$, or AB magnitudes, where $\mu_{int}$ is the intrinsic surface brightness.

\end{document}